\documentclass[
reprint,
superscriptaddress,
eprint,
twocolumn,
amsmath,
amssymb,
prx,
floatfix,
]{revtex4-2}
\usepackage[english]{babel}

\usepackage{graphicx}
\usepackage{dcolumn}
\usepackage{bm}
\usepackage{dsfont}
\usepackage[colorlinks]{hyperref}
\hypersetup{%
    plainpages=true,
    breaklinks=true,
    hypertexnames=false,
    pageanchor=true,
    colorlinks=true,
    linkcolor={blue},
    citecolor={blue},
    urlcolor={blue},
    anchorcolor={black}
}
\usepackage{hhline}
\usepackage{braket, siunitx}
\usepackage{ifthen}
\newboolean{showcomments}
\setboolean{showcomments}{true}

\usepackage{cleveref}
\usepackage{xcolor}
\usepackage{bbding} 
\usepackage{comment}
\usepackage{xspace}
\usepackage{hyperref}
\usepackage{soul}

\crefname{equation}{Eq.}{Eqs.}
\Crefname{equation}{Equation}{Equations}
\crefname{figure}{Fig.}{Figs.} 
\Crefname{figure}{Figure}{Figures}
\crefname{section}{Sect.}{Sects.}
\Crefname{section}{Section}{Sections}
\crefname{table}{Table}{Tables}
\crefname{appsec}{}{Appendices} 

\newcommand{\mynote}[3]{%
  \ifthenelse{\boolean{showcomments}}{%
   \fbox{\bfseries\sffamily\scriptsize#1}%
   {\small$\blacktriangleright$\textsf{\emph{\color{#3}{#2}}}$\blacktriangleleft$}}%
  {%
   \@bsphack
   \@esphack
  }%
}

\newcommand{\EC}{E_\mathrm{C}}
\newcommand{\EJ}{E_\mathrm{J}}
\newcommand{\EL}{E_\mathrm{L}}

\newcommand{\Aphi}{A_\mathrm{\phi}}

\newcommand{\qceff}{Q_\mathrm{C}^{\mathrm{eff}}}
\newcommand{\figlbl}{Fig.}

\newcommand{\panel}[1]{({#1})}
\newcommand{\eqlbl}{Eq.}

\newcommand*{\figref}[2]{%
  \hyperref[{#1}]{%
    ~\ref*{#1}%
    \ifx\\#2\\%
    \else
      \panel{#2}%
    \fi
  }%
}

\begin{document}
\preprint{APS/123-QED}
\title{Characterization and Comparison of Energy Relaxation in Fluxonium Qubits}

\def\LLaffil{Lincoln Laboratory, Massachusetts Institute of Technology, Lexington, MA 02421, USA}
\def\RLEaffil{Research Laboratory of Electronics, Massachusetts Institute of Technology, Cambridge, MA 02139, USA}
\def\Physaffil{Department of Physics, Massachusetts Institute of Technology, Cambridge, MA 02139, USA}
\def\EECSaffil{Department of Electrical Engineering and Computer Science, Massachusetts Institute of Technology, Cambridge, MA 02139, USA}
\def\affilGoogle{\textit{Google Quantum AI, Santa Barbara, CA, USA}}
\def\affilGoogl{\textit{Google Quantum AI, Cambridge, MA, USA}}
\def\equalA{These authors contributed equally}

\author{Kate~Azar}\email{kazar@mit.edu}\affiliation{\LLaffil}\affiliation{\RLEaffil}\affiliation{\EECSaffil}
\author{Lamia~Ateshian}\altaffiliation[Present address: ]{\affilGoogle}\affiliation{\RLEaffil}\affiliation{\EECSaffil}
\author{Mallika~T.~Randeria}\affiliation{\LLaffil}
\author{Ren\'{e}e~DePencier~Pi\~{n}ero}\affiliation{\LLaffil}
\author{Jeffrey~M.~Gertler}\affiliation{\LLaffil}
\author{Junyoung~An}\affiliation{\RLEaffil}\affiliation{\EECSaffil}
\author{Felipe~Contipelli}\affiliation{\LLaffil}
\author{Leon~Ding}\altaffiliation[Present address: ]{\affilGoogl}\affiliation{\RLEaffil}\affiliation{\Physaffil}
\author{Michael~Gingras}\affiliation{\LLaffil}
\author{Kevin~Grossklaus}\affiliation{\LLaffil}
\author{Max~Hays}\affiliation{\RLEaffil}
\author{Thomas~M.~Hazard}\affiliation{\LLaffil}
\author{Junghyun~Kim}\affiliation{\RLEaffil}\affiliation{\EECSaffil}
\author{Bethany~M.~Niedzielski}\affiliation{\LLaffil}
\author{Hannah~Stickler}\affiliation{\LLaffil}
\author{Kunal~L.~Tiwari}\affiliation{\LLaffil}
\author{Helin~Zhang}\affiliation{\RLEaffil}
\author{Jeffrey~A.~Grover}\affiliation{\RLEaffil}
\author{Jonilyn~L.~Yoder}\affiliation{\LLaffil}
\author{Mollie~E.~Schwartz}\affiliation{\LLaffil}
\author{William~D.~Oliver}\affiliation{\RLEaffil}\affiliation{\Physaffil}\affiliation{\EECSaffil}
\author{Kyle~Serniak}\email{kyle.serniak@ll.mit.edu}\affiliation{\LLaffil}\affiliation{\RLEaffil}

\date{\today}

\begin{abstract}
Fluxonium superconducting qubits have demonstrated long coherence times and high single- and two-qubit gate fidelities, making them a favorable building block for superconducting quantum processors. 
We investigate the dominant limitations to fluxonium qubit energy relaxation time $T_1$ using a set of eight planar, aluminum-on-silicon qubits.
We find that a circuit-based model for capacitive dielectric loss best captures the frequency dependence of $T_1$, which we analyze within both a two-level and a six-level energy relaxation model.
We convert the measured $T_1$ into an effective capacitive quality factor $\qceff$ to compare qubits on equal footing, accounting for independently estimated contributions from $1/f$ flux noise and radiative loss to the control and readout circuitry.
We apply this methodology to compare qubits from two fabrication processes: a baseline process and one that applies a fluorine-based wet treatment prior to Josephson junction deposition.
We resolve a small improvement of~(13.8~$\pm$~8.4$)\%$ in the process mean~$\qceff$, indicating that the fluorine treatment may have reduced loss from the metal-substrate interface, but did not address the primary source of loss in these fluxonium qubits.
\end{abstract}

\maketitle
\section{Introduction}
Superconducting qubits are a promising platform for future quantum computing applications~\cite{Kjaergaard_2020, Blais_CQED_2021}.
Most recent advances toward utility-scale superconducting processors have been based on the transmon qubit~\cite{Koch_2007, Arute_2019, Acharya_2023, Acharya_2025}, a weakly anharmonic $LC$ oscillator realized by a Josephson junction (JJ) inductor. 
Though demonstrations with transmon qubits have been promising, alternative circuits with intrinsic resilience to noise and higher fidelity of operations are also being pursued.
The fluxonium qubit~\cite{Manucharyan_2009} is one such alternative, a circuit comprising a capacitor, JJ, and linear inductor in parallel. 
Fluxonium-based architectures have been used to realize millisecond coherence times~\cite{Pop_2014, Somoroff_2023, Ding_2023}, fast single-qubit gates with fidelities exceeding 0.9999~\cite{Ding_2023, Zhang_2021, Rower_2024}, and two-qubit gates with fidelities above 0.999~\cite{Ding_2023, Zhang2024, lin202424daysstablecnotgatefluxonium}.

Identifying dominant decoherence channels is key to improving superconducting qubit coherence and is a clear avenue toward higher gates fidelities. 
Over the past decade, studies have indicated that fluxonium coherence times are limited by a variety of mechanisms including dielectric loss, likely from charge-coupled environmental two level systems~(TLS)~\cite{Nguyen_2019, Hazard_2019, Sun_2023, Ardati_2024, ateshian2025temperaturemagnetic, zhuang2025nonmarkovianrelaxationspectroscopyfluxonium}; $1/f$ flux noise~\cite{Sun_2023, ateshian2025temperaturemagnetic}; and quasiparticles~\cite{Pop_2014}; depending on device parameters and the materials used to construct the qubits.
In particular, the materials and processing associated with JJ fabrication~\cite{Martinis2005_TLS, Oliver2013MaterialsIS, Dunsworth_2017, Mamin_2021, Bilmes_2022, gingras2025improvingtransmonqubitperformance} are expected to play a strong role in the performance of fluxoniums with JJ-array-based superinductors~\cite{Masluk2012, Bell2012}.
Leveraging data from multiple qubits in conjunction with modifications to materials processing can validate process improvements and elucidate process reproducibility, for which consistency will be key for building utility-scale quantum processors~\cite{Yan_2016FluxQbRevisited, Nguyen_2019}.
Qubit-design and materials-process studies have primarily used transmon qubits~\cite{ Place2021newmaterial, Altoe_2022, gingras2025improvingtransmonqubitperformance, bland20252dtransmonslifetimescoherence, murthy2025Fermiidentifyingmaterialslevelsourcesperformance}, with fewer examples utilizing multiple fluxonium qubits~\cite{Nguyen_2019, Gao2025materialsdisorder, Wang2025Fluxonium}.

\begin{figure*}[h!t]
    \centering
    \includegraphics{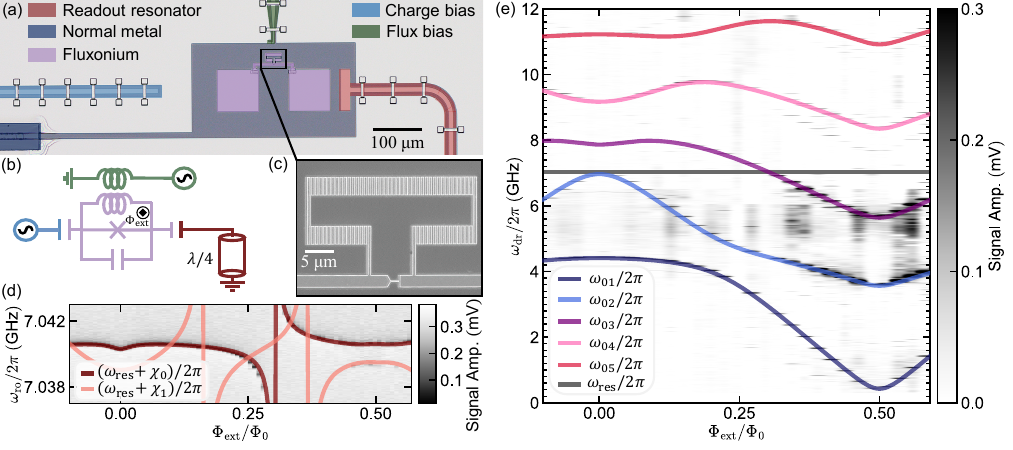}
    \caption{Device design summary. (a) Optical micrograph of a representative fluxonium circuit used in this study, with key elements  highlighted in color. (b) Circuit schematic with colors corresponding to those in (a). (c) Scanning electron micrograph of the qubit JJs from (b). (d) Dressed resonator frequency versus applied magnetic flux bias for qubit $B1$. The dressed readout resonator frequency for the qubit in ground $\omega_{\mathrm{res}} + \chi_0$ and first excited state $\omega_{\mathrm{res}} + \chi_1$ are overlaid. (e) Qubit drive frequency $\omega_{\mathrm{dr}}/2\pi$ versus applied magnetic flux bias for qubit $B1$. Overlaid transitions out of the qubit ground state are highlighted in color, with bare resonator frequency $\omega_{\mathrm{res}}/2\pi$ in dark gray.}
    \label{Fig1}
\end{figure*}

\indent In this Article, we investigate the frequency dependence of the energy relaxation time $T_1$ of eight fluxonium qubits.
To identify the dominant source of loss, we model the effects of various energy-relaxation mechanisms, accommodating for population-transfer dynamics outside of the computational basis of the qubit~\cite{ateshian2025temperaturemagnetic}.
Over a majority of the tunable frequency range, we find the trend in $T_1$ to be well described by a circuit model for capacitive dielectric loss. 
Accounting for the contributions from $1/f$ flux noise and radiative loss to the microwave environment, we transform $T_1$ into an effective capacitive dielectric quality factor $\qceff$.
We observe frequency-dependent variation of $\qceff$ consistent with limitation from TLS defects, and with our experimental protocol determining the mean $\qceff$ to within approximately $\pm20\%$.
We compare the mean $\qceff$ of qubits fabricated with two distinct fabrication processes (here $A$, $B$), used for transmon qubits in Ref.~\cite{gingras2025improvingtransmonqubitperformance} (there ``POR" and ``E"). 
Process $B$ introduces a pad etch substrate cleaning step (described in detail in Sec.~\ref{Dev Desg and Fab}), shown in \cite{gingras2025improvingtransmonqubitperformance} to improve the metal-substrate interface underneath the JJs of transmons resulting in a 2$\times$ improvement in their quality factor.
Here, we resolve a statistically significant difference in the $\qceff$ means of (13.8~$\pm$~8.4)$\%$, indicating that the metal-substrate interface was improved, but was not the predominant interface contributing to the energy relaxation rate in these fluxoniums.
The methodology and analysis used here is extensible to comparisons of any superconducting qubit designs or fabrication processes, the optimization of which is critical for the development of more performant superconducting quantum processors.

\section{Device Design and Fabrication}
\label{Dev Desg and Fab}
The fluxonium circuit~\cite{Manucharyan_2009} is modeled by the Hamiltonian
\begin{equation}
\label{eq: flux Hamiltonian}
    \hat{H} = 4 \EC\hat{n}^2 - \EJ \cos \hat{\varphi} + \frac{1}{2} \EL\left(\hat{\varphi} - 2\pi\frac{\Phi_\mathrm{ext}}{\Phi_{0}} \right)^2,
\end{equation}
which contains an energy term for each of the elements in the circuit.
Here, $\EJ = I_{\mathrm{C}}\Phi_0/(2\pi)$ is the Josephson energy of the JJ, where $\Phi_0=h/(2e)$ is the superconducting magnetic flux quantum and $I_{\mathrm{C}}$ is the JJ critical current, which depends on its size.
$\EC = e^2/(2C_\Sigma)$ is the single-electron charging energy, where $C_\Sigma$ is the total capacitance shunting the JJ. 
Finally, $\EL =  \Phi_0^2/(4\pi^2L_{\mathrm{A}})$ is the inductive energy, where in our devices $L_{\mathrm{A}}$ is the inductance of the JJ array that we use to construct this element.
The operator $\hat{\varphi}$ represents the phase drop across JJ, and is canonically conjugate to the reduced charge $\hat{n}$.
By threading an externally applied magnetic flux $\Phi_{\mathrm{ext}}$ through the loop formed by the JJ and linear inductor, the eigenenergies, eigenstates, and transition matrix elements of the circuit are varied. 

\renewcommand{\arraystretch}{1.5}
\begin{table*}
\centering
\begin{tabular}{c||c|c|c|c|c|c|c|c|c|c|c|c|c} 
Process & Design & $\EJ/h$ & $\EC/h$ & $\EL/h$ & $\omega_{\mathrm{res}}/2\pi$ & $g/2\pi$ & $\kappa/2\pi$ & $\chi_{01}/2\pi$ & $\omega_{01}/2\pi$ & $\sqrt{A_\Phi}$ & $T_1$ & $T_{2}^{\mathrm{R}}$ & $T_{2}^{\mathrm{E}}$ \\
&  & (GHz) & (GHz) & (GHz) & (GHz) & (MHz) & (MHz) & (MHz)& (GHz) & ($\mu\Phi_0/\sqrt{\mathrm{Hz}}$) & ($\mu$s) & ($\mu$s) & ($\mu$s) \\
\hline
$A$ & 1 & 3.54 & 1.05 & 0.53 & 7.090 & 124 & 0.25 & 1.44& 0.362 & 10.4 & $124\ ^{150}_{95}$ & $46\ ^{55}_{33}$ & $112\ ^{133}_{87}$ \\
\hline
$A$ & 2 & 3.94 & 1.04 & 0.53 & 7.182 & 124 & 0.17 &  1.63 & 0.279 & 5.3 & $157\ ^{226}_{117}$ & $77\ ^{91}_{65}$ & $144\ ^{149}_{129}$\\
\hline
$A$ & 3 &  4.38 & 1.02 &  0.53 & 7.273 &  123 & 0.34 & 1.80 & 0.214 & 4.3  & $138\ ^{188}_{92}$ & $76\ ^{86}_{69}$ & $97\ ^{117}_{87}$ \\
\hline
$A$ & 4 & 5.41 & 1.00 &  0.53 & 7.297 & 124 & 0.27 & 2.29& 0.115 & 6.4 & $404\ ^{446}_{343}$ & $107\ ^{153}_{83}$ & $154\ ^{197}_{136}$ \\
\hline 
$A$ & 5 &  7.11 &  0.95 & 0.53 & 7.500 & 120 & 0.60 & 2.20& 0.042 & 2.4 & $129\ ^{152}_{113}$ & $43\ ^{62}_{18}$ & $66\ ^{86}_{39}$ \\
\hline
$B$ & 1 & 3.15 & 1.04 & 0.50 & 7.039 & 118 & 0.29 & 1.11 & 0.427 & 5.2 & $85\ ^{309}_{24}$ & $54\ ^{92}_{40}$ & $86\ ^{154}_{63}$\\
\hline
$B$ & 2 & 3.52 & 1.04 & 0.51 & 7.126 & 120 & 0.30 &  1.30& 0.343 & 5.7 & $101\ ^{118}_{87}$ & $98\ ^{111}_{75}$ & $103\ ^{124}_{89}$  \\
\hline
$B$ & 3 & 3.81 & 1.03 & 0.50 & 7.229 & 120 & 0.35 & 1.39 & 0.284 & 4.3 & $325\ ^{370}_{295}$ & $120\ ^{133}_{95}$ & $137\ ^{142}_{128}$\\
\end{tabular}
\caption{Summary of measured device parameters. Hamiltonian energies, the bare resonator frequency $\omega_{\mathrm{res}}/2\pi$, and qubit-resonator coupling $g/2\pi$ are extracted from measurements of single- and two-tone spectroscopy a function of $\Phi_\mathrm{ext}$ [Fig.~\ref{Fig1}(d, e)]. The linewidth of the readout resonator $\kappa/2\pi$, the relative dispersive shift $\chi_{01}/2\pi$ of the qubit on the resonator, qubit frequency $\omega_{01}/2\pi$, as well as the mean, maximum, and minimum of five repeated measurements of $T_1$, $T_2^{\mathrm{R}}$, and $T_2^{\mathrm{E}}$ are reported at $\Phi_\mathrm{ext}/\Phi_0 = 0.5$. Values are reported as $\text{mean}^{\text{max}}_\text{min}$. $\sqrt{A_\Phi}$ is the extracted flux noise amplitude at 1 Hz.
\label{tab:Fluxonium Device Summary}}
\end{table*}

In order to differentiate qubit designs and the two fabrication processes, we label each qubit with a letter and a number (e.g., ``$A1$").
The letter in the fluxonium labels denotes one of two different fabrication processes.  
All qubits in this work are fabricated using an aluminum-based process on 50~mm intrinsic silicon wafers.
In both process versions, an Al film is deposited on a pre-cleaned wafer and patterned to form all device components: microwave control lines, readout resonators, and qubit shunt capacitors. 
The qubit JJs are deposited via shadow evaporation, using a Ge-based mask patterned for Dolan-bridge style Al-AlOx-Al JJs~\cite{Dolan_1977}.
We fabricate the single JJ branch and an array of 151 series JJs [\figlbl\figref{Fig1}{c}] this way, where the series JJs are used as the fluxonium inductor, which we approximate as a linear inductance.

Our fabrication processes were introduced in ~Ref.~\cite{gingras2025improvingtransmonqubitperformance}, which in part motivated this experiment.
In process $A$, referred to as the process-of-record (POR) in Ref.~\cite{gingras2025improvingtransmonqubitperformance}, the surface of the exposed Si and patterned Al base metallization is cleaned before JJ deposition with an Ar ion mill to remove surface oxides and organic contaminants. 
This recipe was shown to leave Ge-based dry-etch byproducts on the Si surface under the JJ bridge, which were interpreted to be limiting the performance of transmon qubits made using this process.
In process $B$ (labeled ``E" in Ref.~\cite{gingras2025improvingtransmonqubitperformance}), an additional fluorine-based wet etch (commercially available Transene Silox Vapox III) was added before the milling step to clean the substrate without removing the Al base metallization. 
This process was shown to reduce the presence of Ge byproducts and other surface residues.
This process was found to provide a 2$\times$ improvement in the transmon qubit quality factor compared to the process $A$. 

The number in the qubit label denotes a distinct single JJ size used to set the qubit $\EJ$.
The size and number of the array JJs used for the linear inductance, the design of coplanar shunt capacitor pads, charge and flux control lines, and the geometry of the ground plane were identical for all qubits.
As such, all qubits were designed to have nominally the same $\EC$ and $\EL$~(Table \ref{tab:Fluxonium Device Summary}), though $\EC$ decreases slightly with increasing $\EJ$ (corresponding to an increased JJ size and therefore increased parallel-plate capacitance of the JJ).
Each qubit is capacitively coupled to its own readout resonator~[Fig.~\ref{Fig1}(a,b)], which in turn is inductively coupled to a shared microwave feedline. 

The qubits characterized in this study are from three unique chips, two fabricated on the same wafer with process $A$ (one chip with $A1$, $A2$, and $A3$, the other with $A4$ and $A5$) and one fabricated on a separate wafer with process $B$.
The design of qubit $A5$ was identical to the device reported in Ref.~\cite{ateshian2025temperaturemagnetic}, and all qubits share similarities with the qubit designs in Ref.~\cite{randeria2024CQPS}.
Both references utilize devices fabricated using process $A$, albeit drawing from different wafers and fabrication runs.

\section{Qubit Characterizations}
\label{Qubit Characterizations}
We utilize standard microwave control and dispersive-readout techniques (see Appendix~\ref{App: Experimental Setup} for details).
For all measurements, we apply flux bias using a global coil as opposed to the on-chip line, which we cap with a 50 $\Omega$ termination at the mixing chamber stage of the dilution refrigerator [Appendix~\ref{App: Experimental Setup}].
All devices include a TiPt patch labeled ``Normal metal" in ~\figlbl\figref{Fig1}{a}, which we use to prevent screening currents in the ground plane induced by the magnetic field bias.

\begin{figure}[t]
    \centering
    \includegraphics{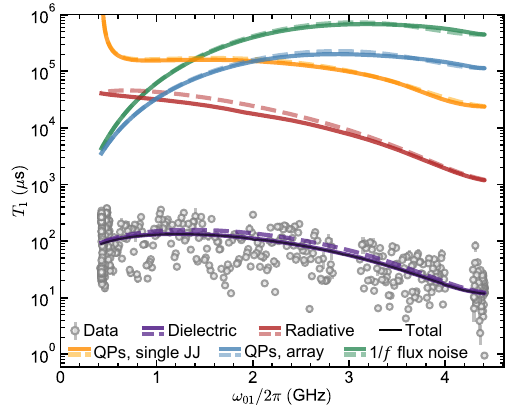}
    \caption{Measured $T_1$ (gray points) versus $\omega_{01}/2\pi$ for qubit $B1$, with predicted $T_1$ due to individual noise models (see Fig.~\ref{AppFig: All T1s, coherence model overlay} for all qubits). Solid lines show predictions for a six-level description of the loss model. Dashed lines show the predictions for a two-level description of the loss model. $1/f$ flux noise and radiative losses are calculated using the measured qubit parameters (Table~\ref{tab:Fluxonium Device Summary}). The dielectric loss model assumes $\qceff$ of $3.11\times 10^5$, with $\epsilon = 0.25$. The quasiparticle models assume $x_{\mathrm{QP}}$ of $1\times 10^{-9}$. The total line represents the six-level description of $T_1$ due to all included noise sources. All models assume an effective qubit temperature of 40 mK and an effective resonator temperature of 65 mK. Gaps in the data near 3, 4.2 GHz are due to insufficient readout contrast [Appendix~\ref{App: Device Characterization}].}
    \label{Fig2}
\end{figure}

We obtain the resonator and qubit parameters by fitting single- and two-tone spectroscopy [\figlbl\figref{Fig1}{d,e}] measurements versus applied magnetic flux bias to extract the bare resonator frequency $\omega_{\mathrm{res}}/2\pi$, qubit-resonator coupling rate $g$, and qubit energy scales $\EJ$, $\EC$, and $\EL$ [Table~\ref{tab:Fluxonium Device Summary}, Appendix~\ref{App: Device Characterization}].
We also report the resonator linewidth $\kappa/2\pi$, relative dispersive shift of the qubit on the resonator $\chi_{01}/2\pi$, and qubit transition frequency $\omega_{01}/2\pi$ at $\Phi_{\mathrm{ext}}/\Phi_0 = 0.5$ [Table~\ref{tab:Fluxonium Device Summary}].
We characterize $T_1$ as a function of $\Phi_\mathrm{ext}$ for each qubit [\figlbl~\ref{Fig2} and Appendix~\ref{App: DC: T1}].
Around $\Phi_{\mathrm{ext}}/\Phi_0 = 0.5$, we performed interleaved measurements of $T_1$, the Ramsey dephasing time $T_2^{\mathrm{R}}$, and the echo dephasing time $T_2^{\mathrm{E}}$ [Appendices~\ref{App: DC: T1},~\ref{App: DC: T2}]. 
The average and standard deviation of all coherence measurements taken at $\Phi_{\mathrm{ext}}/\Phi_0 = 0.5$ are summarized in Table~\ref{tab:Fluxonium Device Summary}.
We use the measurements of $T_1$ and $T_2^{\mathrm{E}}$ taken away from $\Phi_{\mathrm{ext}}/\Phi_0 = 0.5$, where we assume the qubit dephasing to be limited by $1/f$ flux noise, to extract the flux noise amplitude $\sqrt{A_\Phi}$ [Table~\ref{tab:Fluxonium Device Summary}].
We solve the echo pure dephasing rate $1/T^{\mathrm{E}}_\phi~=~|\partial\omega_{01}/\partial\Phi_{\mathrm{ext}}|\sqrt{A_\Phi\mathrm{ln}(2)}$ for $\sqrt{A_\Phi}$, where $|\partial\omega_{01}/\partial\Phi_{\mathrm{ext}}|$ is the slope of the qubit transition frequency as a function of applied magnetic flux bias~\cite{Braumuller2020SQUIDgeometry}.
We observe a wide range of~$\sqrt{A_\Phi}$ for these nominally identical designs, which we are able to partially attribute to noise from our instrumentation [Appendix~\ref{App: DC: Aphi}].

\section{Modeling Energy Relaxation}
\label{Modeling T1}
To identify the loss limiting the measured $T_1$, we consider various noise sources modeled perturbatively using Fermi's golden rule and compare to our data.
The relaxation rate $1/T_1$ is $\sum_\lambda\Gamma_{1}^{\lambda}$, with $\lambda$ indexing the noise sources [Appendix~\ref{App: T1 modeling}].
We consider an extension of usual models, allowing for population dynamics to exist beyond the first two qubit energy levels, a consideration recently shown to be relevant for fluxonium~\cite{ateshian2025temperaturemagnetic}[Appendix ~\ref{App: Nlevel modeling}].
We solve for the time-dependent qubit population vector $\vec{p}(\tau)$ under the differential equation $\frac{d}{d \tau}\vec{p}(\tau) = \boldsymbol{B}\vec{p}(\tau)$, where the transition matrix is
\begin{equation}
        \boldsymbol{B} = \begin{pmatrix}
-\sum\limits_{i\neq1}^{N}\Gamma_{0\rightarrow i} & \Gamma_{1\rightarrow 0} & \cdots & \Gamma_{N\rightarrow 0}\\
\Gamma_{0\rightarrow 1} & -\sum\limits_{i\neq1}^{N} \Gamma_{1\rightarrow i} & \cdots & \vdots  \\
\vdots & \vdots & \ddots & \Gamma_{N \rightarrow N-1} \\
\Gamma_{0\rightarrow N} & \Gamma_{1\rightarrow N} & \cdots & -\sum\limits_{i \neq N}^{N}\Gamma_{N\rightarrow i}
\end{pmatrix},
\end{equation}
and elements $\Gamma_{i\rightarrow j}$ are the transition rates from state $\ket{i}$ to state $\ket{j}$. 
The solution to the differential equation is $\vec{p}(\tau) = \mathrm{exp}({{\boldsymbol{B}}\tau})\vec{p}(0)$.
We fit the decay of the population within the first six levels of the fluxonium back to thermal equilibrium to an exponential to obtain the relaxation time [Appendix~\ref{App: Nlevel modeling}].

To model our data, we consider energy relaxation from capacitively-coupled dielectric loss, $1/f$ flux noise, quasiparticles near the single JJ and within the JJ array, and radiative loss to the control and readout circuitry.
Representative curves for both a traditional two-level and the six-level description of each considered loss model are shown in Fig.~\ref{Fig2}.
We find the model for dielectric loss best describes our data over a majority of the frequency range [Appendix~\ref{Validate T_1 Limitation: All qubits}]. 
For this mechanism, we utilize a phenomenological circuit model that can account for loss caused by imperfect materials, interface oxides, and the average behavior of environmental TLS. 
For this we consider Johnson-Nyquist voltage noise across a resistance in parallel to the qubit capacitance, giving rise to the energy relaxation rate~\cite{NyquistNoise, CaldeiraLeggett}
\begin{equation}
\label{eq: capfgr}
    \Gamma_1^{\mathrm{C}} = \frac{16 \EC}{\hbar Q_{\mathrm{C}}}|\langle 0|\hat{n}|1\rangle|^2\mathrm{coth}\left(\frac{\hbar\omega_{01}}{2k_{\mathrm{B}}T}\right),
\end{equation}
where $Q_\mathrm{C}$ is $1/\mathrm{tan}(\delta)$, and $\mathrm{tan}(\delta)$ is the materials loss tangent [Appendix~\ref{App: T1 Cap Diel. Loss}].
Here, we include a phenomenological frequency dependence of the quality factor~\cite{Wang2015SurfacePartcipationandDielectric},
\begin{equation}
\label{eq: Freq dep Qc}
    Q_{\mathrm{C}} = Q_{\mathrm{C}}^{\prime}\left(\frac{6\times10^9}{\omega_{ij}/2\pi}\right)^\epsilon.
\end{equation}
We find this modification describes the data better than a frequency-independent model [Appendix~\ref{App: Data Processing: Freq Dep}], similar to other literature~\cite{Nguyen_2019, Ardati_2024, ateshian2025temperaturemagnetic}.
As we will describe later but preview here for clarity, to compare devices we ultimately use a value  $\qceff$, which represents the quality factor for a frequency bin-averaged relaxation rate, accounting for contributions from radiative loss to the control circuitry and $1/f$ flux noise. 

\begin{figure*}[t!h]
    \centering
    \includegraphics{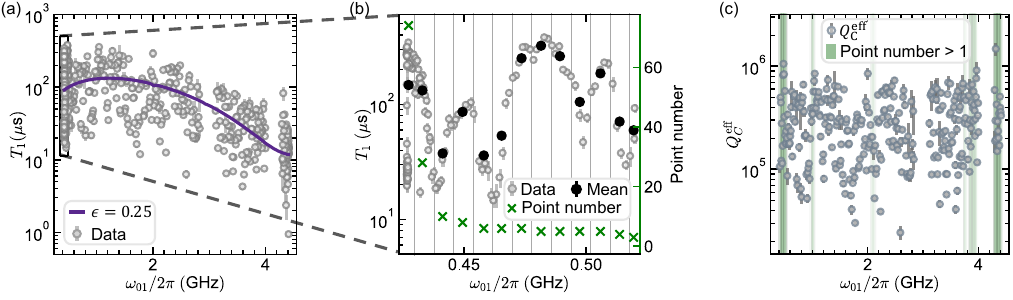}
\caption{Transformation of $T_1$ into $\qceff$ for qubit $B1$. (a) $T_1$ versus qubit transition frequency, with curve for our multilevel model for capacitive loss, using the quality factor as defined by Eq.~(\ref{eq: Freq dep Qc}) with $\epsilon = 0.25$ and $Q_{\mathrm{C}}^{\prime} = 3.11 \times 10^5$, the qubit distribution mean discussed in Section~\ref{sec:Results}. (b) Zoom-in of boxed data in panel (a) where $T_1$ was more finely sampled, additionally visualizing the binned averaging applied to the dataset. Vertical lines indicate the boundaries of adjacent bins, with the right side y-axis corresponding to the number of points within each bin. (c) Extracted $\qceff$ versus qubit transition frequency for the binned $T_1$ values. Green shading indicates frequencies where the number of points per bin was greater than one.}
\label{Fig3}
\end{figure*}

In general the prediction of the six-level description of $T_1$ processes is similar to the two-level description, with the exception of Purcell loss. 
We favor the use of the better motivated, multi-level model in all subsequent analysis.
For all models we assume an effective qubit temperature of 40 mK~\cite{Geerlings_2013, Jin_2015} and an effective resonator temperature of 65 mK~\cite{Sears2012, Yan_2013}. 

\section{Data Processing}
\label{Data Processing}
We observe TLS signatures in the frequency dependence of $T_1$ and a general trend consistent with the capacitive dielectric loss model.
This motivates us to primarily use this model to quantify qubit performance.
In service of this, we first account for contributions to the $T_1$ from our models for $1/f$ flux noise and radiative loss to the microwave environment~[Appendix~\ref{app: spontaneous emission}], using our independent characterizations of each qubit described in Sec.~\ref{Qubit Characterizations} [Appendix~\ref{Validate T_1 Limitation: All qubits}].
We exclude data for which the combined prediction from $1/f$ flux noise and radiative losses is greater than 10$\%$ of the measured decay rate.

We sample $T_1$ more finely in frequency near the $\Phi_{\mathrm{ext}}/\Phi_0=0$ and $\Phi_{\mathrm{ext}}/\Phi_0=0.5$ biases. 
In these regions, we generally observe fine structure, which we attribute to TLSs [Fig.~\ref{Fig3}(b)].
To mitigate bias associated with potential oversampling of TLSs, we apply a binned average to the dataset of 8~MHz width corresponding to the approximate average full-width-at-half maximum of the observed TLS features. 
The frequency spacing between $T_1$ samples is greater than the bin width for a majority of the frequency tuning range~[\figlbl\figref{Fig3}{c} and Appendix~\ref{App: Data Processing: Binning}]. 
With this analysis procedure, we do not distinguish TLS from any background sources of capacitive dielectric loss. 

To account for the $T_1$ dependence on Hamiltonian parameters, we choose to then recast our data into $\qceff$, the quality factor corresponding to the bin-averaged $T_1$ (as opposed to $Q_{\text{C}}^{\prime}$ corresponding to an individual measurement of $T_1$)~[\figlbl\figref{Fig3}{c}, Appendix~\ref{App: Data Processing: Calc Qceff}].
We additionally accommodate for the effect the resonator Lamb shift due to qubit population in higher levels in our model (an effect not included on the for the $T_1$ curves in Figs.~\ref{Fig2} or~\ref{Fig3}), which we detail further in Appendix~\ref{App: Nlevel modeling: energy relax}.
Here, the mean, median, and distribution of measured $\qceff$ serve as metrics that can be used to compare qubits and fabrication processes inclusive of TLS effects.
Though $\qceff$ could in principle have different frequency dependence across qubits and fabrication processes, we use the value of $\epsilon$ that best describes the combined dataset of eight qubits, $\epsilon = 0.25$, in all analysis for consistency [Appendix~\ref{App: Data Processing: Freq Dep}].

\section{Results}
\label{sec:Results}
\begin{figure}[t!]
    \centering
    \includegraphics{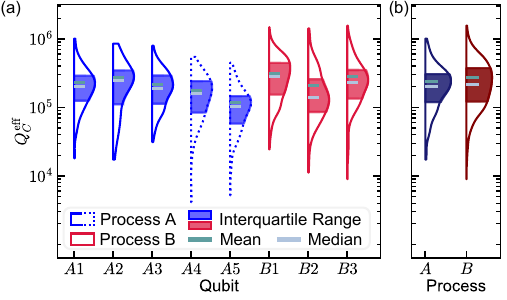}
    \caption{Extracted $\qceff$ distributions. (a) Individual qubit $\qceff$ distributions, where the area of each distribution is normalized to show the percent of counts at each $\qceff$. Blue color corresponds to qubits fabricated with process $A$; red corresponds to process $B$. Shaded regions within individual distributions represent the interquartile range of the dataset. Overlaid teal and gray lines mark the mean and median of each distribution, respectively. (b) Combined distributions of qubits $A1$, $A2$, and $A3$ labeled as Process $A$ (dark blue), and $B1$, $B2$, and $B3$ (Process $B$, dark red) are used to compare like designs across fabrication processes. Dashed outlines in (a) indicate exclusion from the combined distributions shown in panel (b).
    }
    \label{Fig4}
\end{figure}

We compare the distributions of $\qceff$ for each qubit, plotted in Fig.~\ref{Fig4}(a).
The distributions contain largely overlapping $\qceff$ values, and are non-Gaussian distributed.
Though the binning protocol reduces point-to-point correlations caused by TLSs, we do still observe some structure in the binned $T_1$ [\figlbl\figref{Fig3}{b}].
We attribute the uniqueness of the distributions and approximately two order of magnitude range in the $\qceff$ values to the unique spectra and finite number of TLSs that couple to each qubit, which we are not able to fully mitigate in this analysis.
We therefore do not expect these distributions to be identical, which limits the applicability of many statistical tests of similarity. 
We report the mean, median, and standard deviation of $\qceff$ for each qubit in Table~\ref{tab:Mean, Median, CIs}. 
While the median is practically useful for predicting characteristic qubit performance, we choose to apply an interpretable statistical test, Welch's t-test~\cite{Welch1947ttest}, to compare the means between qubits and fabrication processes.
We justify use of this test based on applicability of the central limit theorem to bounded distributions and the assumption that our binning procedure sufficiently reduces point-to-point correlations in the data.

\begin{table}[h]
\begin{tabular}{c||c|c|c|c|} 
Qubit & Mean $\qceff$ & Median $\qceff$ & Std. $\qceff$, $\sigma$ & 95$\%$ CI\\
\hline
$A1$ & 2.32$\times 10^5$ & 1.98$\times 10^5$ &  1.59$\times 10^5$ & $\pm$ $16\%$\\
\hline
$A2$ & 2.68$\times 10^5$ & 2.47$\times 10^5$ & 1.96$\times 10^5$ &  $\pm$ $20\%$\\
\hline
$A3$ &  2.17$\times 10^5$ & 1.91$\times 10^5$ & 1.34$\times 10^5$ &  $\pm$ $16\%$\\
\hline
$A4$ & 1.75$\times 10^5$ & 1.57$\times 10^5$ & 1.13$\times 10^5$ &  $\pm$ $19\%$\\
\hline 
$A5$ &  1.16$\times 10^5$ &  1.03$\times 10^5$ & 0.81$\times 10^5$ &  $\pm$ $17\%$\\
\hline
$B1$ & 3.11$\times 10^5$ & 2.79$\times 10^5$  & 2.03$\times 10^5$ & $\pm$ $10\%$\\
\hline
$B2$ & 2.10$\times 10^5$ & 1.41$\times 10^5$  & 2.06$\times 10^5$ & $\pm$ $20\%$ \\
\hline
$B3$ & 2.81$\times 10^5$ & 2.31$\times 10^5$  &  2.14$\times 10^5$ & $\pm$ $15\%$\\
\hline
$A$ & 2.37$\times10^5$ & 2.00$\times 10^5$  &  1.65$\times 10^5$ & $\pm$ $10\%$\\
\hline
$B$ & 2.75$\times10^5$ & 2.19$\times 10^5$  &  2.11$\times 10^5$ & $\pm$ $8\%$\\
\end{tabular}
 \caption{Summary for individual and combined process $\qceff$ distributions. Means, medians, and standard deviations $\sigma$ of each distribution are reported. We also report the 95$\%$ confidence interval (CI) of the mean as returned by Welch's t-test for each distribution as compared to itself, expressed as a percent of the distribution mean.
 \label{tab:Mean, Median, CIs}}
\end{table}

We use Welch's t-test to calculate the 95$\%$ self-confidence interval [Appendix \ref{App: Stats Tests}] for each distribution, which
bounds the sensitivity of the test: we cannot on average resolve differences in the mean smaller than approximately $\pm$17$\%$ of an individual qubit's mean [Table~\ref{tab:Mean, Median, CIs}].
To compare qubits we compute the pairwise confidence interval [see Appendix~\ref{App: Stats Tests}, Table~\ref{tab: CIs 95}].
We find some distributions appear to be similar and some to be statistically different.
Comparing designs within the same fabrication process (qubits $A1$, $A2$, $A3$, $A4$, and $A5$), a weak trend in the mean as a function of $\EJ$ emerges across otherwise highly overlapped distributions. 
This could possibly be attributed to the increased capacitance from the small junction as its size was increased to vary $\EJ$~[Appendix \ref{App: N level JJ Participation Ratio}]. 
Based on observed TLS densities and effective quality factors of typical shadow-evaporated Al/AlOx/Al JJs~\cite{Martinis2005_TLS,Mamin_2021} this trend is expected; however, further experiments are needed to definitively quantify the small JJ contribution to dielectric loss in our fluxonium qubits.

To compare fabrication processes, we combine the $\qceff$ distributions of $A1$, $A2$, and $A3$ and separately those of the identically designed qubits $B1$, $B2$, and $B3$ [Fig.~\ref{Fig4}, Table \ref{tab: CIs 95}].
We exclude $A4$ and $A5$ from this comparison, as we did not test qubits from process $B$ targeting the same values of $\EJ$.
We find these two composite distributions to be statistically different, with a 95$\%$ confidence interval indicating that the mean of distribution of combined $B$ qubits is higher by $(13.8~\pm~8.4)\%$ of the mean of the corresponding combined $A$ qubits [Appendix~\ref{App: Stats Tests}, Table~\ref{tab: CIs 95 process compare}].
This indicates that the pad etch treatment of process $B$ did slightly improve the effective materials quality of our fluxoniums compared to process $A$ (presumably from removing residues at the metal-substrate interface).
However, the large overlap between the process $\qceff$ distributions and the small difference in the distribution means suggests the metal-substrate interface underneath the JJs is not the primary contributor to dissipation in these qubits.
In Appendix~\ref{app: N level compare with 0 epsilon}, we verify that consideration of a frequency-independent $Q_C$ does not qualitatively change this result.
In Appendix~\ref{App: 2 level modeling}, we show that this result is also preserved when considering the traditional two-level model of qubit energy relaxation.

\section{Conclusions and Outlook}
We have examined the frequency dependence of $T_1$ for eight planar, aluminum-on-silicon, fluxonium qubits sampling from two distinct fabrication processes.
We employ a multilevel model of the energy relaxation dynamics, and find the frequency dependence of $T_1$ is well modeled by a combination of capacitive dielectric loss, radiative loss to the microwave environment, and $1/f$ flux noise. 
Using these three loss models and accounting for the frequency dependence of the qubit transition matrix elements, we convert $T_1$ into a capacitive quality factor $\qceff$ to compare qubits.

Our results indicate that microwave loss associated with the metal-substrate interface underneath the fluxonium JJs do not play as limiting a role in energy relaxation as it did for transmons using the same materials processing~\cite{gingras2025improvingtransmonqubitperformance}. 
We hypothesize that the many JJs comprising the linear inductance of the fluxonium play a role in the reduction of relative $\qceff$ (approximately $3.48\times10^5$ for these fluxoniums as opposed to approximately $6.6\times10^6$ for the transmons in Ref.~\cite{gingras2025improvingtransmonqubitperformance}, both applying a frequency independent capacitive loss model~[see Appendix~\ref{app: N level compare with 0 epsilon}]). 
Taken together, these results point to the metal-air interface or the JJ barrier itself as the main host of fluxonium-limiting TLSs. 
Additional studies seeking to understand the microscopic nature of TLSs will be key to future mitigation strategies and improving superconducting qubit performance.

While our characterization technique does not probe the microscopic properties of the TLSs directly, it consolidates their effects into simple metrics---the $\qceff$ distribution means and 95$\%$ confidence intervals---with which we compare qubits.
Combining data from three qubits per process, we find the 95$\%$ confidence interval comparing the composite distributions to themselves is on the order of~$\pm10\%$ or less of their respective means.
Measuring more qubits would further reduce our estimation error, at the expense of additional measurement time.
This low estimation error indicates that our protocol is sensitive to relatively small changes in performance and can be used in future experiments as an effective probe of new materials, fabrication processes, and qubit designs.

\section{Acknowledgments}
The authors would like to thank A. Almanakly, R. Assouly, W. P. Banner, J. Cummings, A. Di Paolo, A. L. Emser, G. Holland, A. Kou, J. F. Marques, X. Miloshi, and S. K. Tolpygo for valuable assistance and technical discussions. 
This research was funded in part under Air Force Contract No. FA8702-15-D-0001.
L. A. acknowledges support from the NSF Graduate Research Fellowship and from the Laboratory for Physical Sciences Doc Bedard Fellowship. 
M. H. is supported by an appointment to the Intelligence Community Postdoctoral Research Fellowship Program at the Massachusetts Institute of Technology administered by Oak Ridge Institute for Science and Education (ORISE) through an interagency agreement between the U.S. Department of Energy and the Office of the Director of National Intelligence (ODNI). 
J. A. and J. K. acknowledge support from the Korea Foundation for Advances Studies (KFAS).
The views and conclusions contained herein are those of the authors and should not be interpreted as necessarily representing the official policies or endorsements, either expressed or implied, of the U.S. Government. 

\section*{Appendices}

\section{Experimental Setup}
\label{App: Experimental Setup}
\begin{figure}[t]
    \centering
    \includegraphics{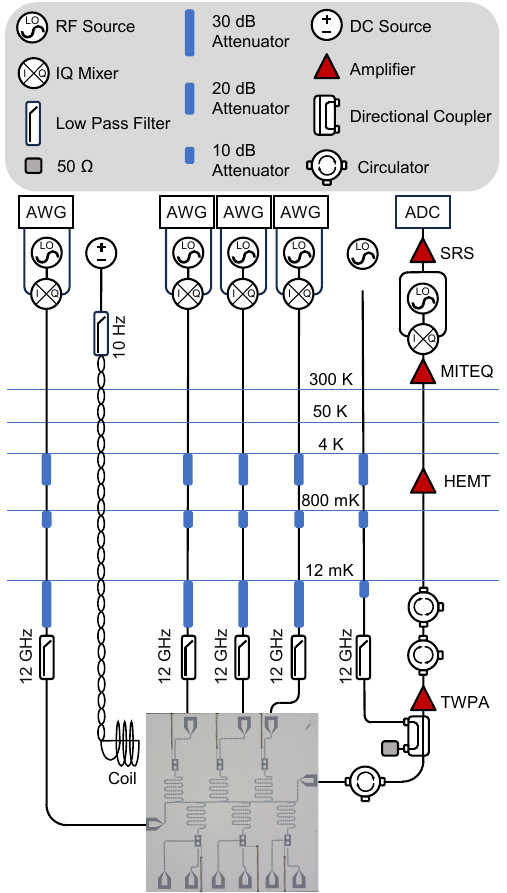}
    \caption{Wiring diagram of experimental setup. Unused qubit control lines (here on the lower side of the sample chip) are connected to the 12 mK stage and capped with 50 $\Omega$ terminations, not pictured.}
    \label{AppFig: Wiring Diagram}
\end{figure}
\begin{figure*}[th!]
    \centering
    \includegraphics{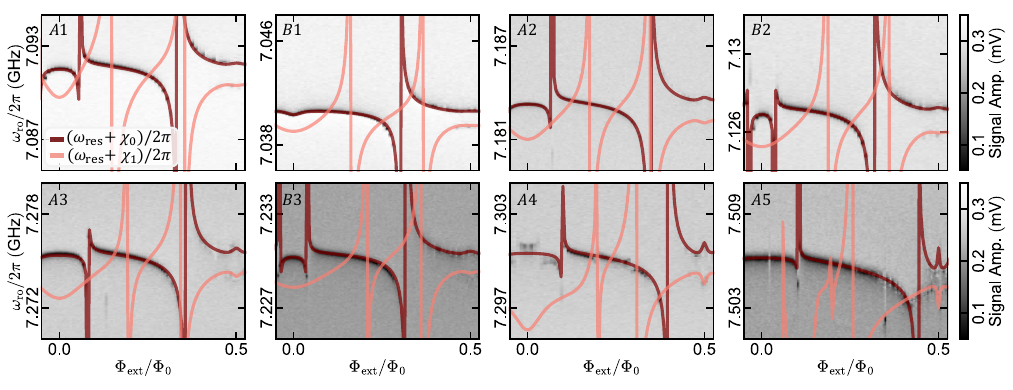}
    \caption{Readout resonator spectroscopy as a function of applied magnetic flux bias, with theory lines for the sum of bare resonator frequency and dispersive shifts $(\omega_{\mathrm{res}} + \chi_0)/2\pi$ and $(\omega_{\mathrm{res}} + \chi_1)/2\pi$ overlaid. Here, we vary the y-axis across plots to accommodate the unique resonator frequencies, while the x-axis is shared. The color bar is also shared across all plots.}
    \label{AppFig: Resonator spectrum}
\end{figure*}
This experiment was performed in a BlueFors XLD1000 dilution refrigerator used to cool the qubits to a temperature of approximately 12 mK. 
Devices were wirebonded into gold-plated copper packages and mounted to a cold finger connected to the mixing chamber stage of the cryostat.
The samples are shielded with three layers of magnetic shielding at the mixing chamber stage.
The innermost shield is made of aluminum, the middle shield of tin-plated copper, and the outer shield is made of Ammuneal A4K.

The dc flux bias is supplied by a coil mounted to the lid of the package, current biased by a Yokogowa GS200 operating in voltage mode, filtered at room temperature by a homemade $10$ Hz low-pass RC filter with a $10~\mathrm{k}\Omega$ series resistance. 
Twisted pair lines connect the dc bias to the sample coil. 
On-chip flux bias lines were present but not used in this experiment.
All unused bias lines (charge and flux) were capped with $50~\Omega$ terminations away from the qubit package, at the mixing chamber stage.

For microwave qubit control, signals were generated by single-sideband mixing of Keysight PXIe M3202A 1 GSa/s arbitrary waveform generators with a Rohde and Schwarz SGS100A SGMA RF source. 
Readout signals were generated by single-sideband mixing of PXIe Keysight M3202A 1 GSa/s arbitrary waveform generators with a Keysight MXG microwave signal generator. 
All ac control and readout signals were attenuated with 60 dB of attenuation distributed throughout the fridge temperature stages, and filtered with a 12 GHz low-pass filter [Fig.~\ref{AppFig: Wiring Diagram}]. 
All control electronics are synchronized with a 10 MHz SRS Rb clock. 

Output signals were amplified first by a traveling wave parametric amplifier (TWPA)~\cite{Macklin_2015} made by MIT Lincoln Laboratory, pumped by a Rohde and Schwarz SGS100A SGMA RF source.
Signals were further amplified by a LNF high electron mobility transistor (HEMT) amplifier at the 4 K stage, and again at room temperature with a MITEQ amplifier, and finally by Stanford Research Systems SR445A amplifier, before being detected by a Keysight M3102A digitizer.

\section{Device Characterization}
\label{App: Device Characterization}
We model the coupled qubit-resonator system with the Hamiltonian, $\hat{H} = \hat{H}_\mathrm{qubit} + \hat{H}_\mathrm{res} + \hat{H}_\mathrm{int}$.
Here, $\hat{H}_\mathrm{qubit}$ is given by Eq.~\ref{eq: flux Hamiltonian}, the bare resonator Hamiltonian is
\begin{equation}
\hat{H}_\mathrm{res} = \hbar\omega_{\mathrm{res}}(\hat{a}^{\dagger}\hat{a}+1/2),
\end{equation}
and the interaction Hamiltonian between the qubit and resonator is
\begin{equation}
\hat{H}_\mathrm{int}=\hbar g \hat{n}(\hat{a} + \hat{a}^{\dagger}).
\end{equation}

The coupling constant $g=\omega_{\mathrm{res}}\frac{C_c}{C_\Sigma}\sqrt{\pi Z_0/R_q}$, where the characteristic impedance of the readout resonator $Z_0=50~\Omega$, and $R_q = h/(2e)^2$ is the resistance quantum.
The frequency shift of the resonator due to the qubit occupying state $\ket{i}$ (otherwise referred to as the dispersive shift or Lamb shift), is then given by~\cite{Maucharyan_phaseslips2012}
\begin{equation}
    \chi_{i} = \sum_{i\neq j}\frac{2g^2|\langle i|\hat{n}|j\rangle|^2\omega_{ij}}{\omega_{ij}^2 - \omega_{\mathrm{res}}^2}
\label{eq: dispersive shift}
\end{equation}
which is the basis for the dispersive measurement scheme employed in this work.

\begin{figure*}[h!t]
    \centering
    \includegraphics{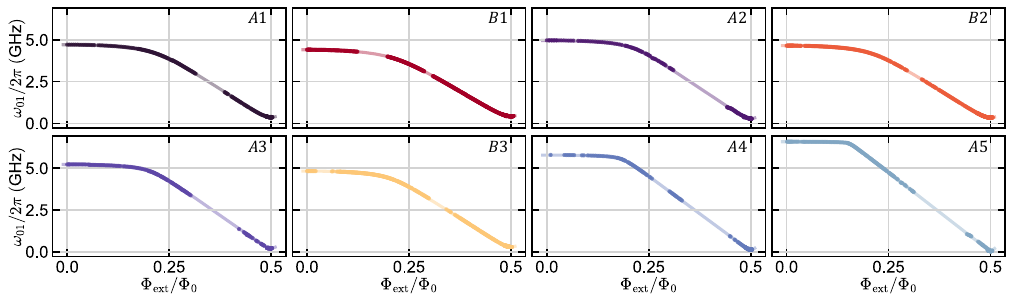}
    \caption{Qubit transition frequency $\omega_{01}/2\pi$ versus applied magnetic flux $\Phi_\mathrm{ext}$. Points indicate measured frequencies, and the line is the calculated frequency from the qubit Hamiltonian parameters listed in Table~\ref{tab:Fluxonium Device Summary}.}
    \label{AppFig: Qubit spectrum}
\end{figure*}

We show the readout response around the resonator frequency versus applied flux bias for all qubits in Fig.~\ref{AppFig: Resonator spectrum}, which we fit to Eq.~\ref{eq: dispersive shift} to extract the bare resonator frequency $\omega_{\mathrm{res}}$ and qubit-resonator coupling rate $g/2\pi$.
We additionally show the measured qubit frequency $\omega_{01}/2\pi$, with the overlay as predicted by each qubit's Hamiltonian parameters in Fig.~\ref{AppFig: Qubit spectrum}.

\subsection{Heralded Measurement}
\label{App: DC: heralding}
When the qubit transition energy $\hbar\omega_{01}$ becomes small relative to the thermal energy scale $k_BT$, the equilibrium population in $\ket{1}$ reduces readout contrast. 
In our experiment, at qubit transition frequencies of $\approx$ 1 GHz and below, a heralded measurement scheme is used for all qubit characterizations \cite{Serniak2018hot}. 
In this scheme, a qubit pre-measurement $m_1$ is performed before the qubit control pulses and is then correlated with a post-measurement $m_2$ to determine the change in qubit state as a result of the control pulses. 
A wait time of 5 $\mu$s ($>5/\kappa$ for all readout resonators) is added between $m_1$ and any qubit control pulses to allow for measurement photons to dissipate.
To determine the results in this heralded scheme, the averaged correlation function of the two measurements $\langle m_1m_2\rangle$ is fit to a function specific to each measurement.

\subsection{\texorpdfstring{$T_1$}{T1} Measurements}
\label{App: DC: T1}
\begin{figure*}[h!t]
    \centering
    \includegraphics{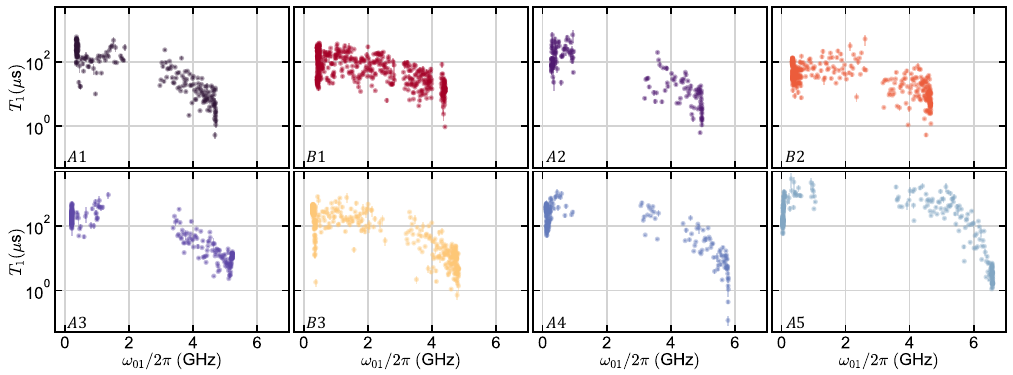}
    \caption{Measured $T_1$ for all qubits as a function of transition frequency $\omega_{01}/2\pi$. Gaps in measured data correspond to regions of insufficient readout contrast (see Fig.~\ref{AppFig: Resonator spectrum}).}
    \label{AppFig: All qubits T1}
\end{figure*}

For each $T_1$ measurement, the qubit and resonator system was characterized at the corresponding flux bias, according to the methods described in Appendix~\ref{App: Device Characterization}. 
By applying a sideband modulated cosine pulse to drive the qubit frequency on the charge bias line, the population in the qubit $|0\rangle$ and $|1\rangle$ states is inverted, after which the population is allowed to freely decay for a variable amount of time prior to measurement. 
The delay times were spaced logarithmically for each $T_1$ trace to capture the wide distribution of decay times. 
The resultant signal decay curve was fit to an exponential function to extract $T_1$, with error bars corresponding to one standard deviation of error as output from our fitting protocol. 
We exclude from our analysis any $T_1$ measurements for which this error bar was greater than twice the value of the $T_1$ time, which indicates a failure of our automated calibration routine. 
The measured $T_1$ versus $\omega_{01}/2\pi$ are shown in Fig. \ref{AppFig: All qubits T1} for all fluxoniums measured.

\subsection{Dephasing Measurements}
\label{App: DC: T2}
At $\Phi_\mathrm{ext}/\Phi_0 = 0.5$, we characterize the Ramsey and echo dephasing times of each qubit, where each measurement utilized a heralding pre-measurement. 
To characterize the Ramsey dephasing, we apply a resonant $\pi/2$-pulse to create an equal superposition state, wait for a variable delay time, and then apply a second resonant $\pi/2$-pulse (with variable phase to induce oscillations, as opposed to a real frequency detuning) before dispersively measuring the qubit state. 
The Echo dephasing time is characterized similarly, adding a $\pi$-pulse centered between the pair of $\pi/2$ pulses. 
The measurements are fit to $\langle m_1(\tau)m_2(\tau)\rangle = A\times\mathrm{exp}(-\tau/T_2^{\{R,E\}})\times\cos(\omega\tau + \phi)$.
The average and standard deviations of repeated measurements of the Ramsey and Echo dephasing times $T_2^{\mathrm{R}}$ and $T_2^{\mathrm{E}}$, are reported in Table~\ref{tab:Fluxonium Device Summary}.

\subsection{Characterizing \texorpdfstring{$A_\phi$}{Aphi}}
\label{App: DC: Aphi}
\begin{figure*}[h!t]
    \centering
    \includegraphics{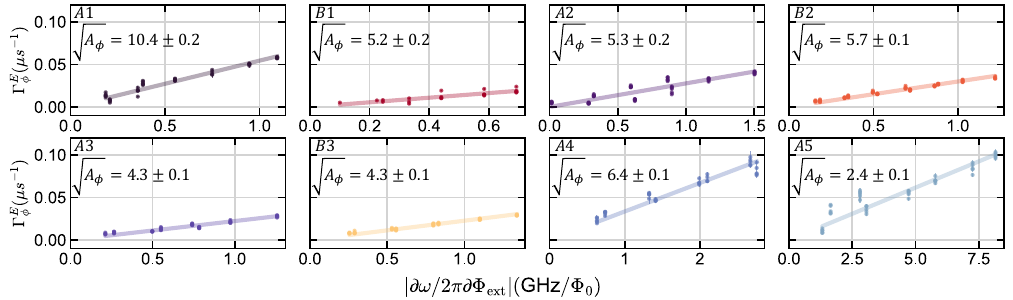}
    \caption{Measured $\Gamma^E_\phi$ versus external flux sensitivity for all qubits with linear fit used to extract $\sqrt{\Aphi}$ for each qubit, which is reported with the standard error on the fit. We note that the y-axes are shared across all plots, however the x-axis varies as was necessary to accommodate each qubit's unique transition spectrum and the range over which $\Gamma^E_\phi$ was characterized.}
    \label{AppFig: All qubits aphi}
\end{figure*}
To characterize the flux noise amplitude of each qubit, repeated interleaved measurements of $T_1$ and $T_\phi^\mathrm{E}$ were taken over a narrow range around $\Phi_{\mathrm{ext}} = 0.5\Phi_0$. 
Away from $\Phi_{\mathrm{ext}} = 0.5\Phi_0$, where we expect the dephasing to be limited by $1/f$ flux noise, the data from the spin-echo experiments were fit with a decay envelope $\langle m_1(\tau)m_2(\tau)\rangle = A\times\mathrm{exp}[-\tau/(2T_1) -(\tau/T_\phi^{\mathrm{E}})^2]\times\cos(\omega\tau + \phi)$.
In this fit function, $T_1$ was constrained to the value obtained from the immediately preceding measurement. 
As shown in Fig~\ref{AppFig: All qubits aphi}, we analyze the dependence of $\Gamma_\phi^\mathrm{E}= 1/T_\phi^\mathrm{E}$ on the absolute value of the qubit spectrum slope in flux, where we expect $\Gamma^{\mathrm{E}}_\phi = |\frac{\partial\omega}{\partial\Phi}|\sqrt{A_\phi \mathrm{ln}(2)}$. 
We fit the data to a linear function to extract $\sqrt{A_\phi}$.
Given that dephasing in this model vanishes at $\Phi_{\mathrm{ext}}/\Phi_0 = 0.5$ where $\frac{\partial\omega}{\partial\Phi} = 0$, the fit enforces an intercept with the origin \cite{Braumuller2020SQUIDgeometry}.
Measurements at $\Phi_{\mathrm{ext}} = 0.5\Phi_0$ where we do not expect dephasing to be limited by first-order flux noise were not included in the linear fit. 

\begin{figure}[h!]
    \centering
    \includegraphics{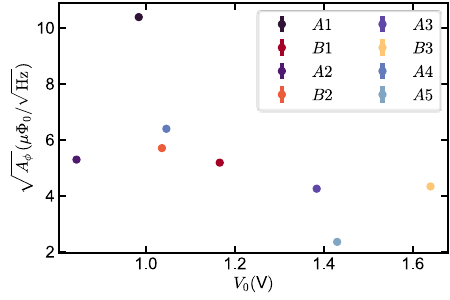}
    \caption{Extracted $\sqrt{A_\phi}$ versus $V_0$, the voltage required to  bias the fluxonium to $\Phi_\mathrm{ext} = \Phi_0$.}
    \label{AppFig: Aphi vs mutual}
\end{figure}
We observe a wide range of $\sqrt{A_\phi}$ across the fluxoniums~[Fig \ref{AppFig: All qubits aphi}]. 
Noise that couples to the qubit from the flux bias coil and from defects in the environment contribute to this flux noise. 
Noise from our instrumentation would appear as correlation between $\sqrt{A_\phi}$ and the effective coupling of the bias line to the qubit (larger coupling corresponds to a lower applied voltage to achieve one $\Phi_0$ of applied bias). 
Fig.~\ref{AppFig: Aphi vs mutual} shows the flux noise amplitude $A_\phi$ versus the voltage periodicity for each qubit. 
We observe a weak correlation between the voltage periodicity and the extracted flux noise amplitude, indicating that global bias noise is not itself sufficient to describe the variation in $\sqrt{A_\phi}$. 

\section{Modeling \texorpdfstring{$T_1$}{T1}}
\label{App: T1 modeling}
Fermi's golden rule is used to model the transition rate between two qubit energy eigenstates $\ket{i}$, $\ket{j}$, under the influence of a perturbatively treated noise source \cite{Dirac_1927}, which is given by
\begin{equation}
    \Gamma_{\lambda} = \frac{2}{\hbar^2}|\langle i|\hat{D}_\lambda|j\rangle|^2 S_\lambda(\omega_{ij}).
\end{equation}
Here, $S_\lambda(\omega_{ij}) = \frac{1}{2}\int_{-\infty}^{\infty}\langle \lambda(0)\lambda(t) + \lambda(t)\lambda(0)\rangle e^{-i\omega_{ij} t}dt$, which is the symmetrized, two-sided power spectrum of the time-fluctuating noise $\lambda(t)$, and $\hbar\omega_{ij}$ is the transition energy between the eigenstates $\ket{i}, \ket{j}$. 
The coupling of this noise to the qubit system is generically described by a linear interaction term in the Hamiltonian of the form $\hat{D}_\lambda \lambda(t)$.
The qubit susceptibility to different noise sources is contained within the transition matrix element $|\langle i|\hat{D}_\lambda|j\rangle|^2$.

\subsection{Circuit Models of Loss}
\label{App: T1 modeling: Circuits}
For circuit-based models of dissipation, we model noise as a fictitious resistor in the qubit circuit. 
Under this class of models, the power spectrum is obtained by considering either voltage fluctuations across a resistor in parallel to the qubit circuit, or current fluctuations through a resistor in series with a qubit circuit element. 
Following the Caldeira-Leggett model, the symmetrized noise spectra for Johnson-Nyquist voltage and current fluctuations are, respectively \cite{NyquistNoise, CaldeiraLeggett, DevoretLesHouches}, 
\begin{equation}
    S_V(\omega_{ij}) = \hbar \omega_{ij} \mathrm{Re}[Z(\omega_{ij})]\mathrm{coth}\left(\frac{\hbar\omega_{ij}}{2k_BT}\right),
\end{equation}
and
\begin{equation}
    S_I(\omega_{ij}) = \hbar \omega_{ij} \mathrm{Re}[Y(\omega_{ij})]\mathrm{coth}\left(\frac{\hbar \omega_{ij}}{2k_BT}\right),
\end{equation}
where $Z(\omega_{ij})$ is the impedance of our effective circuit and $Y(\omega_{ij}) = 1/Z(\omega_{ij})$ is the admittance. 

\subsubsection{Capacitive Dielectric Loss}
\label{App: T1 Cap Diel. Loss}
Lossy dielectric materials that couple to the electric field of the qubit via the operator $2e\hat{n}$ are modeled in this work as Johnson-Nyquist voltage noise across the qubit capacitor. 
The relevant impedance for the voltage noise spectrum is calculated from the parallel capacitive and resistive branches, $Z(\omega_{ij}) = (i\omega_{ij}C_\Sigma + 1/R)^{-1}$. 
All other qubit elements are treated as perfect and lossless in this model.
We substitute the quality factor for a parallel resonance circuit, $Q_\mathrm{C} = \omega_{ij} RC_\Sigma$ \cite{pozar_microwave_2012}, into our impedance expression to obtain $Z(\omega)$ = $\frac{1}{Q_{\mathrm{C}}\omega_{ij} C_\Sigma}$.
Together this gives the energy relaxation rate \cite{SmithThesis},
\begin{equation}
        \Gamma_1^{\mathrm{C}} = \frac{16 \EC}{\hbar Q_{\mathrm{C}}}|\langle 0|\hat{n}|1\rangle|^2\mathrm{coth}\left(\frac{\hbar\omega_{01}}{2k_BT}\right).
\end{equation}
where $Q_\mathrm{C}$ indicates the quality factor of the lossy capacitor. 
Though the construction of this model typically assumes quality factor of the circuit to be frequency independent, we lift this assumption (Eq.~\ref{eq: Freq dep Qc}), which we find better describes our data, similar to other literature~\cite{Wang2015SurfacePartcipationandDielectric, Nguyen_2019, Sun_2023, Ardati_2024, ateshian2025temperaturemagnetic}.
We do not explicitly model the fine frequency structure associated with TLSs, and instead use the frequency dependent, binned average effective quality factor $\qceff$ in our main analysis to characterize and compare qubits.

\subsubsection{Inductive Loss}
Dual to the capacitive loss model is a phenomenological model for lossy dielectric materials that couple to the magnetic field of the qubit through the operator $\phi_0\hat{\varphi}$, where $\phi_0 = \Phi_0/2\pi$ are modeled in this work as Johnson-Nyquist current noise through a series resistive element. 
In this work we instead consider of two microscopically motivated models --- $1/f$ flux noise and QP tunneling through JJ array --- instead of this phenomenological inductive loss model. 
For more details on this inductive circuit loss model, see Refs.~\cite{SmithThesis, Hazard_2019, Zhang2024, ateshian2025temperaturemagnetic} and references therein.


\subsubsection{Spontaneous Emission to Control and Readout Circuits}
\label{app: spontaneous emission}
We utilize circuit-based models for loss through the charge and flux control lines, as well as for resonance-enhanced (Purcell) emission through the readout resonator, described in detail here.

\textbf{Charge control line:}
We model radiative loss to the charge control line as voltage noise from a 50 $\Omega$ environment coupled to the qubit across the capacitance between the control line and the qubit, $C_\mathrm{d}$.
The relaxation rate is written as \cite{manucharyanThesis, Zhang_2021, ateshian2025temperaturemagnetic} 
\begin{equation}
    \Gamma_1^{\mathrm{cd}} = \frac{16\EC C_{\mathrm{d}}^2}{\hbar^2e^2}|\langle 0 |\hat{n}| 1\rangle|^2\mathrm{coth}\left(\frac{\hbar \omega_{01}}{2k_{\mathrm{B}}T}\right).
\end{equation}
From finite element simulations, $C_{\mathrm{d}}$ is estimated to be 20~aF, which is used for all qubits. 
From the qubit Hamiltonian parameters, the coupling capacitance, and assumed qubit effective temperature, we calculate the contribution of this mechanism to all qubit $T_1$ times in our dataset. 
This mechanism is accounted for in the main text comparative analysis, as well as in two-level level analysis detailed in Appendix \ref{App: 2 level modeling}.

\textbf{Flux control line:}
Analogous to the charge control line, we model radiative loss to the flux control line as current noise from a 50 $\Omega$ environment coupled via the mutual inductance between the control line and the qubit, $M_{\mathrm{d}}$. 
Combined with the current noise spectrum, this gives the relaxation rate \cite{ateshian2025temperaturemagnetic}, 
\begin{equation}
    \Gamma_1^{\mathrm{fd}} = \frac{8\pi^2\EL^2 M_{\mathrm{d}}^2}{\hbar^2 \Phi_0^2}|\langle 0 |\hat{\varphi}| 1\rangle|^2\mathrm{coth}\left(\frac{\hbar \omega_{01}}{2k_{\mathrm{B}}T}\right).
\end{equation}
The flux line design and loop area for all qubits in this dataset is identical to that in Ref.~\cite{ateshian2025temperaturemagnetic}, with mutual inductance $M_\mathrm{d} = \Phi_0/21.5$ mA used for all qubits in this dataset \cite{ateshian2025temperaturemagnetic}.
From the qubit Hamiltonian parameters, the mutual inductance, and the assumed qubit effective temperature, we  calculate the contribution of this mechanism to all qubit $T_1$ times in our dataset.
This mechanism is accounted for in the main text comparative analysis, as well as in the N = 2 level analysis detailed in Appendix ~\ref{App: 2 level modeling}.

\textbf{Readout Resonator:}
\begin{figure}[h!]
    \centering
    \includegraphics{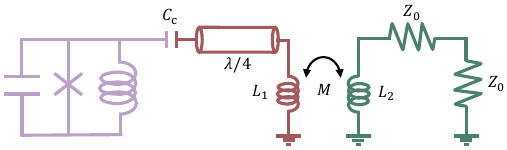}
    \caption{Effective circuit diagram used for calculation of  Purcell loss. Purple elements correspond to the fluxonium qubit circuit, red elements correspond to the readout resonator, and green elements correspond to the microwave feedline.}
    \label{App: Circuit Purcell Loss}
\end{figure}

We utilize a circuit based model for Purcell decay through the readout resonator, in contrast to other models in literature that treat the resonator quantum mechanically \cite{Zhang2024}. 
This was chosen to more accurately model the frequency dependence of the coupling between all qubit levels and the resonator. 
We model the readout resonator as a $\lambda/4$ transmission line with characteristic impedance $Z_0 = 50$ $\Omega$, capacitively coupled on one end to the qubit and inductively coupled to the feedline on the other. 
The feedline is modeled as an infinitely long transmission line via termination with a $50$ $\Omega$ resistor on either end  [Fig. \ref{App: Circuit Purcell Loss}]. 
The relaxation rate due to this mechanism is written as \cite{ateshian2025temperaturemagnetic} 
\begin{equation}
    \Gamma_{1}^{\mathrm{P}} = \frac{8e^2\omega_{01}}{\hbar}\left(\frac{C_{\mathrm{c}}}{C_{\Sigma}}\right)^{2}|\langle 0|\hat{n}|1\rangle|^2  \mathrm{Re}[Z_{\mathrm{in}}(\omega_{01})] \mathrm{coth}\left(\frac{\hbar\omega_{01}}{2k_BT}\right)
\end{equation}
where the coupling capacitance $C_{\mathrm{c}} = \frac{\hbar g C_{\Sigma}}{2e\omega_{\mathrm{res}}}\sqrt{\frac{\pi}{2\hbar Z_0}}$, and $C_\Sigma$ is the total qubit capacitance.
$Z_{\mathrm{in}}(\omega_{01})$ is the input impedance of the lossy environment the of the feedline filtered by the readout resonator, as seen by the qubit.   

For the mode with frequency $\omega_{\mathrm{res}}$, $Z_{\mathrm{in}}$ is written as
\begin{equation}
    Z_{\mathrm{in}}(\omega) = Z_0\frac{\omega^2M^2\cot(\frac{\pi\omega}{2\omega_{\mathrm{res}}}) + 2jZ_{0}^2}{2Z_{0}^2\cot(\frac{\pi\omega}{2\omega_{\mathrm{res}}}) + j\omega^2M^2},
\label{eq: T1 models: Purcell loss input impedance}
\end{equation}
where $M$ is the mutual inductance between the resonator and feedline \cite{pozar_microwave_2012}. 
Assuming this coupling is weak such that $\omega M \ll Z_0$, we rewrite Eq.~\ref{eq: T1 models: Purcell loss input impedance} as
\begin{equation}
    Z_{\mathrm{in}}(\omega) \approx \frac{1}{\frac{\omega_{\mathrm{res}}^2M^2}{Z_0^3} + \frac{2j\pi\omega}{2\omega_{\mathrm{res}}Z_0}} = \frac{1}{1/R + 2j\omega C}.
\label{eq: T1 models: Purcell loss weak coupling}
\end{equation}
where by matching terms with the parallel RLC circuit, we find $R = Z_0^3/\omega_{\mathrm{res}}^2M^2$, and $C = \pi/2\omega_{\mathrm{res}}Z_0$. 
The quality factor is defined as $Q = \omega RC$, which for the obtained values of $R$ and $C$ can be rewritten as
\begin{equation}
    Q = \frac{\pi Z_0^2}{2\omega_{\mathrm{res}}^2M^2}.
\end{equation}
We do not directly measure the mutual inductance between the readout resonator and the feedline, but we do measure the resonator quality factor, $Q_{\mathrm{res}} = \omega_{\mathrm{res}}/\kappa$, which we use as a substitute for the mutual inductance (assuming that the resonator is over-coupled to the feedline),
\begin{equation}
    M = \frac{Z_0}{\omega_{\mathrm{res}}}\sqrt{\frac{\pi}{2Q_{\mathrm{res}}}}.
\end{equation}

For qubits with transition frequencies within 1 GHz of their readout resonator frequencies, we find this model for radiative loss through the readout resonator becomes strong enough to limit the $T_1$ in addition to capacitive dielectric loss. 
Under a six-level model for this mechanism, the limit to $T_1$ is stronger across the qubit tuning range than in a two-level model, with features corresponding to higher qubit transitions crossing the resonator (see solid red curves in Fig.~\ref{AppFig: All T1s, coherence model overlay}).

\subsection{Quasiparticle-Induced loss}
\label{App: T1 modeling: xqp}
Here we model energy relaxation due to nonequilibrium quasiparticles tunneling across the qubit junctions. 
We first assume the energy hierarchy, $\varepsilon_{\mathrm{QP}},\ \hbar\omega_{ij}\ll\Delta$, where $\varepsilon_{\mathrm{QP}}$ is the characteristic energy of the quasiparticles,  $\hbar\omega_{ij}$ is the qubit transition energy, and $\Delta$ is the superconducting gap. 
In this low energy regime, we additionally assume that the qubit energy is larger than the quasiparticle energy, which allows us to simplify the noise spectral density to \cite{Catelani_2011, Pop_2014}
\begin{equation}
    S_{\mathrm{QP}}(\omega_{ij}) = x_{\mathrm{QP}}\frac{8\EJ}{\pi\hbar}\sqrt{\frac{2\Delta}{\hbar\omega_{ij}}}.
\end{equation}
The consequence of this assumption is that quasiparticles can only relax the qubit, and not excite it as they do not have the requisite energy. 
We reintroduce this possibility into the quasiparticle noise spectral density by adding a contribution reduced by a Boltzmann factor $e^{\hbar\omega_{ij}/k_BT}$, consistent with detailed balance.
The loss from quasiparticles tunneling across the junction couples via the $\mathrm{sin}({\hat{\varphi}}/2)$ operator, which combined with the noise spectrum gives the relaxation rate \cite{Catelani_2011, Pop_2014}
\begin{equation}
    \Gamma_1^{\mathrm{QP}} = \frac{16\EJ x_{\mathrm{QP}}}{\pi\hbar}\sqrt{\frac{2\Delta}{\hbar\omega_{01}}}|\langle 0|\mathrm{sin}(\hat{\varphi}/2)|1\rangle|^2 (1 + e^{\hbar\omega_{01}/k_BT}),
\end{equation}
where $x_{\mathrm{QP}}$ is the normalized quasiparticle density. 
For quasiparticles tunneling in the JJ array inductor, we sum the contribution from every JJ, each with corresponding phase drop $\hat\varphi/N_\mathrm{array}$ where $N_\mathrm{array}$ is the total number of JJs in series. 
As $N_\mathrm{array}$ is typically large (in our case 151), we approximate $\sin\hat{\varphi}\approx \hat{\varphi}$ to get the rate,
\begin{equation}
    \Gamma_1^{\mathrm{QP,A}} = \frac{2\EL x_{\mathrm{QP}}}{\pi\hbar}\sqrt{\frac{2\Delta}{\hbar\omega_{01}}}|\langle 0|\hat{\varphi}|1\rangle|^2 (1 + e^{\hbar\omega_{01}/k_BT}).
\end{equation}

Assuming a spatially homogeneous, normalized QP density $x_{\mathrm{QP}}$, a value of $1\times 10^{-6}$ or higher would be required for tunneling through the single JJ branch to describe the measured $T_1$.
This quasiparticle density is significantly larger than reported in recent measurements~\cite{serniak2019,connolly2024}.
The same $x_{\mathrm{QP}}$ predicts a $T_1$ due to QP tunneling in the inductor JJs that underestimates the data at low transition frequencies. 
The array tunneling model can align with the data at low frequencies using an $x_{\mathrm{QP}}= 1\times 10^{-8}$ or less, for which QP tunneling across the single JJ branch would not be limiting.
At these frequencies, we continue to see signatures of dielectric loss limitation (TLS-like features), indicating that limitation from quasiparticles is unlikely. 
Therefore, we assume QP tunneling to be a negligible effect in all analysis.

\subsection{\texorpdfstring{$1/f$}{1/f} flux noise}
\label{App: T1 modeling: 1/f flux}
To model $1/f$ flux noise, we consider fluctuations in flux bias coupling to the fluxonium Hamiltonian [\eqlbl~\ref{eq: flux Hamiltonian}], which couple to the qubit through the inductor via the operator $\phi_0\hat{\varphi}$. 
We write the power spectrum of this noise as
\begin{equation}
    S_\Phi(\omega_{ij}) = \left(\frac{2\pi A_\phi}{\omega_{ij}}\right)^\alpha
\end{equation}
where $A_\phi$ is the noise amplitude referenced at 1 Hz \cite{Yan2012, Yan_2016FluxQbRevisited}. 
The method used to extract $A_\phi$ is described in detail in Appendix ~\ref{App: DC: Aphi}, assuming $\alpha = 1$. 
The energy relaxation rate is written as
\begin{equation}
    \Gamma_1^\Phi = 4\pi\left(\frac{2\pi \EL}{\hbar\Phi_0}\right)^2|\langle 0|\hat{\varphi}|1\rangle|^2\frac{A_\phi}{\omega_{01}}
\end{equation}
where temperature dependence of this model in included implicitly in the value of $A_\phi$ \cite{ateshian2025temperaturemagnetic, Sun_2023}. 

\section{N-Level Energy Relaxation Model}
\label{App: Nlevel modeling}
\subsection{Modeling energy relaxation dynamics}
\label{App: Nlevel modeling: energy relax}
Qubit population transfer through higher energy levels has been shown to be relevant to relaxation of qubits with energy splitting on the order of $k_\mathrm{B}T$ ~\cite{yu2005multiphotonmultilevel,hann2018,ateshian2025temperaturemagnetic}.
We consider the population dynamics using the time evolution of the population vector $\vec{p}$, which tracks six lowest energy levels of the circuit (after which we see the results from our modeling converge, for all qubits).
The vector $\vec{p}$ evolves under the equation $\frac{d}{d \tau}\vec{p}(\tau) = \boldsymbol{B}\vec{p}(\tau)$ \cite{ateshian2025temperaturemagnetic}.
The matrix $\boldsymbol{B}$ is defined as
\begin{equation}
        \boldsymbol{B} = \begin{pmatrix}
-\sum\limits_{i\neq1}^{N}\Gamma_{0\rightarrow i} & \Gamma_{1\rightarrow 0} & \cdots & \Gamma_{N\rightarrow 0}\\
\Gamma_{0\rightarrow 1} & -\sum\limits_{i\neq1}^{N} \Gamma_{1\rightarrow i} & \cdots & \vdots  \\
\vdots & \vdots & \ddots & \Gamma_{N \rightarrow N-1} \\
\Gamma_{0\rightarrow N} & \Gamma_{1\rightarrow N} & \cdots & -\sum\limits_{i \neq N}^{N}\Gamma_{N\rightarrow i}
\end{pmatrix}
\end{equation}
where elements $\Gamma_{i\rightarrow j}$ are transition rates from state $\ket{i}$ to state $\ket{j}$, where at present we are considering the transitions induced by a \textit{single} loss model. 
The solution to the differential equation is $\vec{p}(\tau) = e^{{\boldsymbol{B}}\tau}\vec{p}(0) = \boldsymbol{V}e^{\boldsymbol{S}\tau}\boldsymbol{V}^{-1}\vec{p}(0)$, where $\boldsymbol{B}$ is diagonalized by $\boldsymbol{V}\boldsymbol{S}\boldsymbol{V}^{-1}$.
Here $\boldsymbol{S}$ is a diagonal matrix of real eigenvalues associated with the columns of matrix $\boldsymbol{V}$, which are composed of the eigenvectors of $\boldsymbol{B}$.

The population evolves according to each eigenmode of $\boldsymbol{B}$ as 
\begin{equation}
    \vec{p}(\tau) = c_0\vec{v}_0 + c_1\vec{v}_1 e^{-\gamma_1 \tau} + ... c_N\vec{v}_N e^{-\gamma_N \tau}
    \label{eq: multirate population decay}
\end{equation}
with $\begin{pmatrix}c_0 & c_1 &\cdots c_N  \end{pmatrix}^T = \boldsymbol{V}\vec{p}(0)$, being the coefficients of the initial state in the eigenbasis of $\boldsymbol{B}$.
We use an equilibrium state as the initial state of the system $\vec{p}(0)$ with occupation probabilities given by the Boltzmann distribution, assuming a temperature of 40 mK. 
We invert the population between the first two states and solve for the time evolution of $\vec{p}(\tau)$ back to the equilibrium state. 

With higher energy levels allowed to participate in population dynamics, $T_1$ maintains the definition of being the timescale upon which a perturbed qubit population returns to thermal equilibrium, however these higher levels can introduce interesting effects.
The solution to this model is by definition non-exponential, however our measured decay envelopes do not noticeably deviate from an exponential shape (this is further discussed and quantified in Appendix \ref{AppFig: Quantifying Exponentialness}).
We find for the Hamiltonian parameters of the qubits in this dataset, it is important to consider the multi-rate nature of this model to accurately capture the qubit population dynamics. 
Furthermore, we find a consequential impact on the decay signal from the higher-state Lamb shifts of the readout resonator, which also requires an accounting of all system rates in order to model. 
Due to the range of Hamiltonian parameters and observed impact of the readout resonator on the extracted decay timescale, we follow a protocol distinct from that \cite{ateshian2025temperaturemagnetic}, where the eigenvalue of the mode with greatest overlap with the initial state of the system was chosen as the decay rate. 

In this work, we solve the rate matrix differential equation for $\vec{p}(\tau)$.
Using the extracted qubit Hamiltonian parameters and qubit-resonator coupling, we also calculate the Lamb shifts of the resonator due to qubit population in each considered energy level. 
We find resonator response for each qubit state $\ket{i}$ according to 
\begin{equation}
    S_{21}^i(\omega) = 1 - \frac{Q_{\mathrm{Load}}/Q_{\mathrm{Coup}}}{1+\frac{2iQ_{\mathrm{Load}}\Delta\omega_i}{\omega_{\mathrm{res}}+\chi_i}},
\label{eq: S21}
\end{equation}
where $Q_{\mathrm{Load}}$ and $Q_{\mathrm{Coup}}$ are the loaded and coupling $Q$s of the resonator, respectively, and $\Delta\omega_i$ is the difference between the probe frequency $\omega$ and the dressed frequency of the resonator $\omega_{\mathrm{res}} + \chi_i$.
For this modeling, we make the assumption that $Q_{\mathrm{Load}} = Q_{\mathrm{Coup}} = \omega_{\mathrm{res}}/\kappa$, and that we probed the resonator at frequency $(\omega_\mathrm{res}+\chi_0)/2\pi$

The location in IQ space of each higher state resonator frequency can then be calculated via \eqlbl~\ref{eq: S21}. 
To match experiment, we then rotate the simulated response to maximize signal contrast ($S^i_{21,\mathrm{rot}}$).  
Finally, we calculate an expected $T_1$ decay signal as 
\begin{equation}
    s(\tau) = \left|\sum_i^{N=6}p_i(\tau)\mathrm{Re}\{S^i_{21,\mathrm{rot}}\}\right|,
\end{equation}
where $p_i(\tau)$ is the population in the $i$th state as a function of the $T_1$ measurement delay time, and $S_{21}^i$ is the rotated location of the state in the IQ plane. 
We fit $s(\tau)$ to an exponential to extract the expected decay rate.

To simulate a $T_1$ time informed by multiple loss channels, the architecture outlined previously is extended by summing the rate matrices $\boldsymbol{B}_{\mathrm{tot}} = \sum_m \boldsymbol{B}_m$, $m$ represents each included loss model.
This is analogous to summing the rates due to each loss mechanism in an two-level model, where $\Gamma_1^{\mathrm{tot}} = \sum_m \Gamma_1^m$.
For the main text comparisons, we sum loss matrices for capacitive dielectric loss, $1/f$ flux noise, and radiative losses to the microwave environment to describe the $T_1$ (and calculate $\qceff$) of each qubit in the dataset. 

\subsection{Identifying Limiting $T_1$ Mechanisms}
\label{Validate T_1 Limitation: All qubits}
In Figure~\ref{AppFig: All T1s, coherence model overlay}, we show the measured $T_1$ for all qubits with each coherence model we consider overlaid. 
For the plotted six-level models we do not include the effect of the resonator Lamb shift on the $T_1$, and instead fit the decay of $p_1(t)$ to an exponential.
We use the device specific measured parameters for our models of $1/f$ flux noise and Purcell loss, summarized in Table~\ref{tab:Fluxonium Device Summary}.
We assume an $x_\mathrm{QP}$ of $1\times10^{-9}$ for models of quasiparticle tunneling for each device. 
The dielectric loss model uses the mean $\qceff$ reported in \ref{tab:Mean, Median, CIs} and $\epsilon = 0.25$.
\begin{figure*}[h!t]
    \centering
    \includegraphics{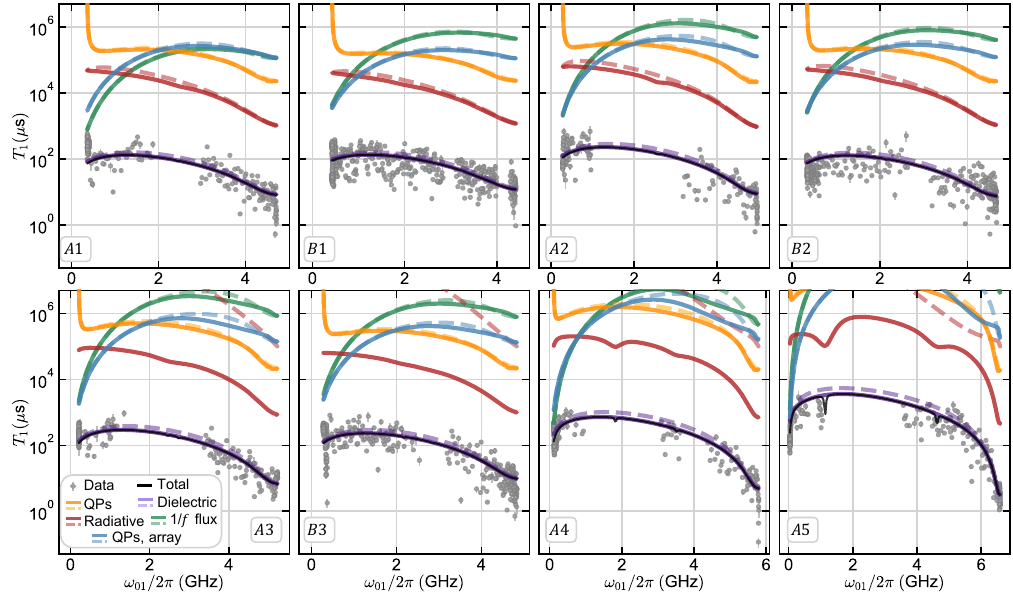}
    \caption{Measured $T_1$ versus $\omega_{01}/2\pi$ for each qubit. Note that each subplot's x axis is varied to accommodate each qubit's frequency tuning range. The y axis range is shared for all plots. Overlay lines show predicted $T_1$ due to different models. Solid lines indicate models for which a six-level treatment is considered, dashed lines indicate a two-level treatment. All models assume an effective qubit temperature of 40 mK and an effective resonator temperature of 65 mK. $1/f$ flux noise, Purcell loss, and dielectric loss curves use qubit specific parameters summarized in Tables~\ref{tab:Fluxonium Device Summary}, \ref{tab:Mean, Median, CIs}. The phenomenological frequency dependence is used for the dielectric loss model, with $\epsilon = 0.25$. Quasiparticle models assume $x_\mathrm{QP} = 1\times10^{-9}$.}
    \label{AppFig: All T1s, coherence model overlay}
\end{figure*}

We see that the capacitive dielectric loss describes the data over a majority of the qubits frequency tuning range. 
For some qubits, limitation from $1/f$ flux noise or Purcell losses are more prevalent in certain regions of the flux range, particularly for qubits $A1$, $A4$, and $A5$. 

\section{Data Processing}
\label{App: Data Processing}
\begin{figure*}[h!t]
    \centering
    \includegraphics{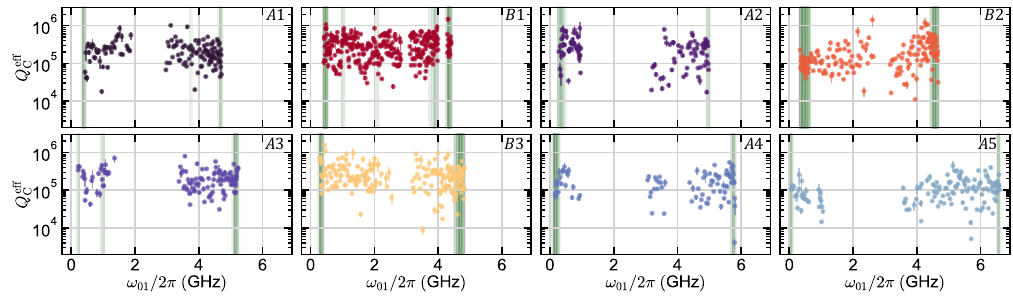}
    \caption{Extracted $\qceff$ data used in the main text, for each qubit, versus qubit transition frequency. The dielectric loss model includes the frequency dependence according to Eq.~\ref{eq: Freq dep Qc} with $\epsilon = 0.25$. Green vertical shaded regions indicate areas where more than one $T_1$ data point was within the same bin. In those regions, the $\qceff$ is calculated from the bin average $T_1$ value.}
    \label{AppFig: N level freq dep Qc}
\end{figure*}

In this section we describe the data processing done to convert the measured $T_1$ times to the $\qceff$ distributions used in the main text for all comparisons. 

\subsection{Binned Average}
\label{App: Data Processing: Binning}
As described in the main text, we apply an 8 MHz binned average to the $T_1$ data across the frequency tuning range of each device. 
The binned average reduces the influence of potential oversampling of specific TLSs on the data analysis by averaging over these features.
The window size of 8 MHz was chosen as the average observed TLS feature full-width-at-half-maximum.

The binning primarily affects data taken near $\Phi_{\mathrm{ext}}/\Phi_0 = 0$ and $0.5$. 
We summarize the number of $T_1$ samples and the number of points remaining after the 8 MHz binned average is applied for each qubit in Table ~\ref{tab:2, N level qubit T1 sample sizes}. 
We see that a majority of the frequency tunable range for each qubit is not impacted by the binning [Fig.~\ref{AppFig: N level freq dep Qc}].

\subsection{Converting to \texorpdfstring{$\qceff$}{qceff}}
\label{App: Data Processing: Calc Qceff}
To extract $\qceff$, we solve the differential equation for population evolution using a rate matrix composed of multiple loss models $\boldsymbol{B}_{\mathrm{tot}} = \sum_m \boldsymbol{B}_m$, where indices $m$ represent each loss model.
At this stage of the analysis, we only include losses in the matrix $\boldsymbol{B}_{\mathrm{tot}}$ due to $1/f$ flux noise, radiative emission to the microwave environment, and capacitive dielectric loss.
We utilize scipy.optimize.minimize \cite{2020SciPy} with the Nelder-Mead method \cite{Nelder1965, Gao2012Implementing} to solve for $\qceff$ from our measured $T_1$.
We define our cost function as squared difference between the predicted $T_1$ from a trial $\qceff$ and the measured $T_1$. 
We exclude data points limited significantly by flux noise and/or radiative losses, defined in this analysis by whether the contributions from those models are greater than 10$\%$ of the measured rate. 
We then assume the remaining data is representative of the capacitive dielectric loss.
The number of points in the final $\qceff$ distribution for each qubit is summarized in ~\ref{tab:2, N level qubit T1 sample sizes}.
Flux noise at low frequencies as well as radiative losses through the readout resonator are both limiting for qubits $A4$ and $A5$ in the dataset, and commensurately they have the fewest number of points remaining after this processing in their $\qceff$ distributions [Fig.~\ref{AppFig: N level freq dep Qc}, Table~\ref{tab:2, N level qubit T1 sample sizes}].

\subsection{Frequency Dependence of the \texorpdfstring{$\qceff$}{qceff} Distributions}
\label{App: Data Processing: Freq Dep}
\begin{figure}[h!]
    \centering
    \includegraphics{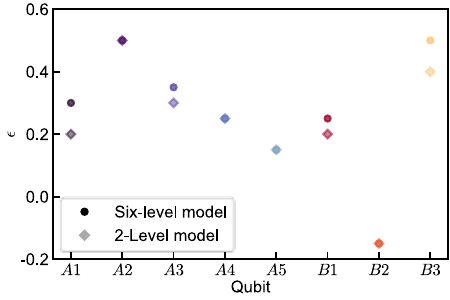}
    \caption{Extracted frequency dependence of the $\qceff$ distributions, $\epsilon$, under the six-level model (circles) and  2-level model (diamonds), for each qubit. }
    \label{AppFig: Device specific epsilons}
\end{figure}

To find the $\epsilon$ in Eq.~\ref{eq: Freq dep Qc} that best describes the entire dataset, we calculate $\qceff$ for each qubit using varying $\epsilon$ from -1 to 1 in steps of 0.05. 
For each qubit at each $\epsilon$, we calculate $\log_{10}(\qceff) - \log_{10}(\overline{\qceff})$, where $\overline{\qceff}$ is the mean of the qubit $\qceff$ distribution. 
We minimize the variance of this parameter combining data from all qubits over the range of tested epsilons to find the best fit, $\epsilon = 0.25$.
We plot the $\qceff$ distributions with this best epsilon versus the qubit transition frequency in Fig.~\ref{AppFig: N level freq dep Qc}.

Though we use the same frequency dependence for each qubit in the main text analysis, we note that each qubit has a unique frequency dependence, varying between $\epsilon = -0.15$ and $\epsilon = 0.50$.
We show the best $\epsilon$ for each qubit for both the six-level analysis and for a two-level analysis in Fig.~\ref{AppFig: Device specific epsilons}.
Most qubits in the dataset show an increase in the average extracted $\qceff$ from high to low transition frequency,  however, this trend is broken by qubit $B2$. 
We note that applying the globally optimized $\epsilon = 0.25$ does not seem to decrease the applicability of this model, given the spread of measured $\qceff$ as a function of qubit frequency.
Though the frequency trend in $\qceff$ is not universal across all devices, by fitting the entire dataset together we are able to compare the distributions on equal footing and mitigate fit bias from qubits with less data at low frequency. 

\section{Validation tests for six-Level Relaxation Model}
Here, we justify the application of exponential decay fit functions to our $T_1$ dataset under the six-level model of population dynamics.

\subsection{Quantifying exponential dynamics}
\begin{figure}[h!t]
    \centering
    \includegraphics{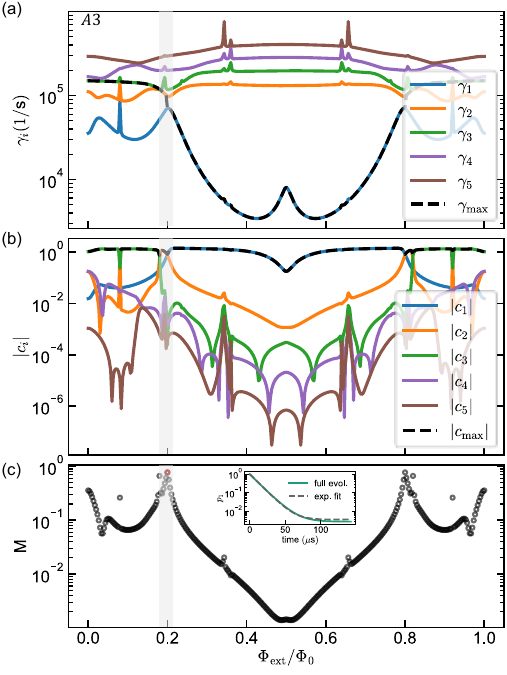}
    \caption{Quantifying exponential behavior under the six-level model for energy relaxation. Plot uses the parameters of the qubit with least exponential behavior (highest M value), $A3$. (a) Decay rates $\gamma_i$ of each rate matrix normal mode versus flux. The rate associated with most dominant mode, $\gamma_{\mathrm{max}}$ is shown with a black dashed line. (b) Overlap with the initial state of each normal mode of the rate matrix. Largest overlap $|c_{\mathrm{max}}|$ shown with black dashed line. (c) M metric versus flux, where high values correspond to less exponential behavior, and low values correspond to more exponential behavior. The maximum M value is highlighted in red. The vertical shaded region corresponds to the least exponential region (this is symmetric in flux, though we only shade the region within $\Phi_{\mathrm{ext}}/\Phi_0 = 0, 0.5$), as determined by the maximum value of M. Inset shows the decay of $p_1$ at the flux with the highest M value, with gray dashed overlay corresponding to a single exponential fit of the decay of $p_1(t)$.}
    \label{AppFig: Quantifying Exponentialness}
\end{figure}
Deviation from exponential decay was not observed in this experiment. 
It is of note, however, that multi-exponential decay dynamics have been observed in other fluxonium coherence studies in literature \cite{Pop_2014, Nguyen_2019}. 
Following Ref.~\cite{ateshian2025temperaturemagnetic}, to quantify our expectations of exponential behavior under a six-level treatment of the population dynamics, we calculate a vector quantity $\vec{\delta}$, defined as
\begin{equation}
    \vec{\delta} = \vec{p}(0) - c_0\vec{v}_0 - c_{k, \mathrm{max}}\vec{v}_{k, \mathrm{max}}.
\end{equation}
Here, we subtract from our equilibrium population vector $\vec{p}(0)$ the contributions of the steady state, or ``stationary mode" $c_0\vec{v}_0$, and the most dominant mode $c_{k, \mathrm{max}}\vec{v}_{k, \mathrm{max}}$. 
We define the eigenvectors to have unit length, such that $||\vec{v}_i|| = 1$.
The most dominant mode is defined as the eigenmode of the rate matrix with the greatest overlap with the initial state of the system, $k,\mathrm{max} = \mathrm{argmax}_{i\neq 0}|c_{i}|^2$, and should be most characteristic of the decay of $\vec{p}(t)$.
Though the main text does not use the decay rate of the most dominant mode as the $T_1$ of the system, we find this method useful to quantify the contribution of all matrix modes to the decay dynamics, which we use to understand the applicability of an exponential fit to the population decay in our system.  

The vector $\vec{\delta}$ contains the population in the nondominant, non-steady state modes of the system. 
The length of $\vec{\delta}$ is
\begin{equation}
    M = \sqrt{\sum_i \delta_i^2},
\end{equation}
which is used as a quantitative metric to benchmark the exponential behavior of our system.
For each device we calculate the decay rates $\gamma_i$, the vector overlaps $|c_i|$, and the length of $\vec{\delta}$, M, across the flux tuning range [Fig. \ref{AppFig: Quantifying Exponentialness}].
We use a rate matrix accommodating for capacitive dielectric loss with the device average $\qceff$, $\epsilon = 0.25$, $1/f$ flux noise, and radiative losses to the microwave environment. 
At the greatest values of M for each device, we examine the decay of $p_1(\tau)$ and compare it to a single exponential decay curve. 

Across all qubits in this study, our model does not predict a significant deviation from exponential decay even at the least exponential (highest M value) flux bias, where we show an exponential fit to the decay of $p_1$ at that bias to confirm [\figlbl\figref{AppFig: Quantifying Exponentialness}{c} inset]. 
This is consistent with our observations of exponential behavior in experiment.

\subsection{Effect of population leakage on dispersive \texorpdfstring{$T_1$}{T1} measurements}
\begin{figure}[t]
    \centering
    \includegraphics{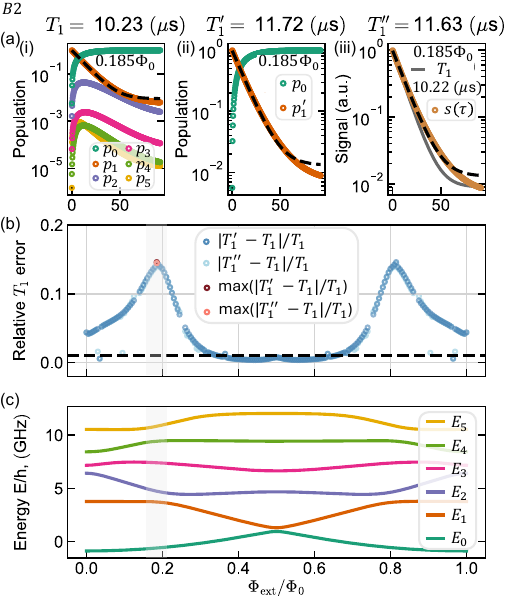}
    \caption{Simulating the effect of state misassignment of higher level populations under different measurement techniques. Calculations shown are for the qubit with highest observed relative errors, $B2$. (a) Comparison of extracted $T_1$ times under different methods. (i) Full population evolution at $\Phi_{\mathrm{ext}}/\Phi_0 = 0.185$, the flux with highest relative error for heralded measurements. Dashed line is an exponential fit to the decay of $p_1$, with the plot title corresponding to the extracted decay timescale. (ii) Decay of $p_0$, $p_1$ with all higher state population assigned to $p_1$ at at $\Phi_{\mathrm{ext}}/\Phi_0 = 0.185$. Plot title corresponding to the extracted decay timescale from an exponential fit to $p_1$. (iii) Decay of signal $s(\tau)$ at the bias with highest relative error for dispersive measurements, $\Phi_{\mathrm{ext}}/\Phi_0 = 0.185$. The dark gray curve (behind the orange points) corresponds to the $T_1$ as extracted from fitting the decay of $p_1$ (where $p1$ is not shown), to compare with the dashed black curve exponential fit of $s(\tau)$. (b) Relative error versus flux for the two measurement techniques. Black dashed line indicates 1$\%$ relative error. (c) Energy spectrum of qubit $B2$ for the first N = 6 levels. The vertical shaded region corresponds to the biases with the highest relative errors, as observed in panel (b). This is symmetric in flux, though we only shade between $\Phi_{\mathrm{ext}}/\Phi_0 = 0, 0.5$.}
    \label{AppFig: Quantifying Msrmt Err}
\end{figure}
Here, we discuss the importance of the inclusion of the Lamb shifts in the simulation of the energy relaxation dynamics. 
We plot the relative error as the $T_1$ extracted from a single exponential fit to $p_1$ (dark gray curve in \figlbl\figref{AppFig: Quantifying Msrmt Err}{a(iii)}), here labeled $T_1$, minus the decay rate as extracted from a single exponential fit to $s(\tau)$ (our simulated signal influenced by the resonator, corresponding to the black dashed curve in \figlbl\figref{AppFig: Quantifying Msrmt Err}{a(iii)})), here $T_1^{\prime\prime}$, divided by $T_1$ [\figlbl\figref{AppFig: Quantifying Msrmt Err}{b}]. 
We see the highest relative error values being between bias values of $\Phi_{\mathrm{ext}}/\Phi_0 = 0$ and $\approx$ 0.2 [\figlbl\figref{AppFig: Quantifying Msrmt Err}{a(iii), b}], with a maximum error of about 15$\%$ observed for qubit $B2$. 
This high relative error indicates that it is indeed necessary to consider the response of the readout resonator for these qubit parameters.
Though qubits with different Hamiltonian parameters may be less sensitive to this effect \cite{ateshian2025temperaturemagnetic}, we choose to apply the same model considerations to all qubits in this dataset.

\subsection{Effect of population leakage on heralded \texorpdfstring{$T_1$}{T1} measurements}
\label{App: N level: heralding}
In this section, we address the potential error in our measured $T_1$ data due to population leakage outside the first two energy levels of the qubit, assuming the data was measured with the heralded scheme described in Appendix ~\ref{App: DC: heralding}.
To model this, we fit an exponential function to the decay $p_1(\tau)$ under two cases: total population misassignment to the $|0\rangle$ state of the qubit, and total population misassignment to the $|1\rangle$ state of the qubit. 

In the case where all higher level population is assigned to the $|0\rangle$ state, the decay of $p_1$ is unaffected and the single exponential fit used to extract $T_1$ is identical to the case of perfect population assignment. 
We therefore compare the values of extracted $T_1$ as obtained from fitting $p_1$ with and without all higher level populations assigned to $p_1$ [\figlbl\figref{AppFig: Quantifying Msrmt Err}{a(ii)}].
We plot the relative error as the $T_1$ extracted from a single exponential fit to $p_1$ with perfect population assignment, here labeled $T_1$, minus the decay rate as extracted from a single exponential fit to $p_1$ with all higher state populations misassigned to it ($T_1^{\prime}$), divided by $T_1$ [\figlbl\figref{AppFig: Quantifying Msrmt Err}{b}]. 
We see a relative error of up to 13$\%$ due to population misassignment in this measurement scheme, with the largest relative errors occurring between $\Phi_{\mathrm{ext}}/\Phi_0$ = 0 and 0.3. 

Heralded measurements were used in this experiment when the qubit transition energy $\hbar\omega_{01}$ became small relative to the thermal energy $k_{\mathrm{B}}T$, around $\omega_{01}/2\pi \approx $ 1 GHz, or $0.7 \gtrapprox \Phi_{\mathrm{ext}}/\Phi_0 \gtrapprox 0.3$ on the qubits in this dataset. 
In that region, the relative error from population misassignment in this scheme is significantly lower, on the order of 1$\%$. 
Though population misassignment could have occurred in this measurement scheme, we conclude that it is likely a small effect on the measured $T_1$ values.
We therefore choose to neglect the impact of this error in our analysis.

\section{Statistical testing}
\label{App: Stats Tests}
We use statistical testing to quantitatively compare the $\qceff$ distributions. 
As our distributions are non-Gaussian distributed, a natural choice would be to use a non-parametric test such as the Kolmogorov-Smirnov (KS) \cite{Kolmogorov1933, smirnov1948table} test or the Mann-Whitney U (MWU) \cite{MannWhitey1947}. 
Crucially, however, the presence of TLS features that will be unique to each qubit's local microscopic environment implies that the underlying distributions of $\qceff$ in the well-sampled limit will never be identical qubit-to-qubit. 
We therefore choose to apply statistical comparison methods that do not rely on the assumption that the all moments of the compared distributions are identical.
This criterion rules out the KS test, which does have this hypothesis. 
The MWU test would still be an appropriate choice, however it loses interpretability for distributions with distinct shapes. 
Given the variation we observe in our distributions, we favor using Welch's two-sided t-test for its interpretability as a comparison of means.

Welch's two-sided t-test is a statistical test comparing the difference in means of two distributions, accommodating for the unique variance of the distributions \cite{Welch1947ttest}. 
The test calculates the test statistic $t_0$, defined as 
\begin{equation}
t_0 = \frac{\bar{x}_1 - \bar{x}_2}{\sqrt{\frac{\sigma_1^2}{n_1} + \frac{\sigma_2^2}{n_2}}},
\label{eq: stats test t-statistic}
\end{equation}
where $\bar{x}_1$ and $\bar{x}_2$ are the means of compared distributions, $\sigma_1^2$ and $\sigma_2^2$ are their variances, and $n_1$ and $n_2$ are the number of their samples.

The test also calculates the degrees of freedom $\nu$, defined in this case as
\begin{equation}
\nu = \frac{(\frac{\sigma_1^2}{n_1}+ \frac{\sigma_2^2}{n_2})^2}{\frac{(\frac{\sigma_1^2}{n_1})^2}{n_1 - 1} + \frac{(\frac{\sigma_2^2}{n_2})^2}{n_2 - 1}}
    \label{eq: stats test df}
\end{equation}
Using the t-statistic and the degrees of freedom, the p-value is calculated from the t-distribution probability density function (PDF), given by
\begin{equation}
f(t, \nu) = \frac{\Gamma(\frac{\nu + 1}{2})}{\sqrt{\pi\nu}\Gamma(\frac{\nu}{2})}(1 + \frac{t^2}{\nu})^{-(\nu + 1)/2},
\label{eq: t-dist PDF}
\end{equation}
\begin{figure}
    \centering
    \includegraphics[width=0.95\linewidth]{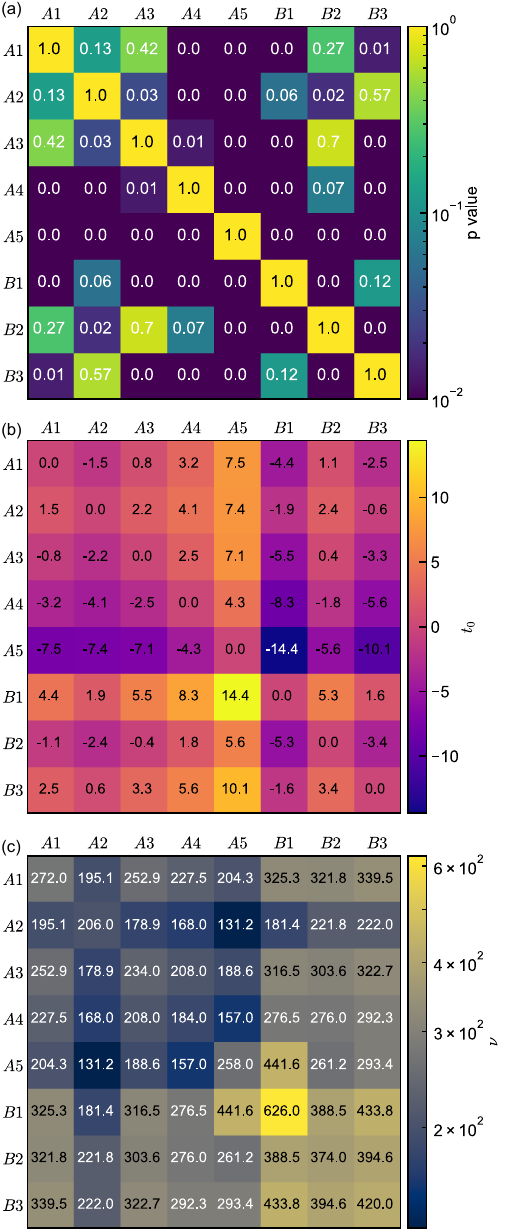}
    \caption{Supplemental parameters from Welch's t-test for pairwise comparisons of all qubit $\qceff$ distributions in the main text [Fig.~\ref{Fig4}]. (a) P-values for all pairwise comparisons. For a 95$\%$ threshold, a p-value of 0.05 or greater indicates similarity in mean according to this test. Values lower than 0.05 indicates that this test finds the distribution means to be distinct. Values less than 0.01 are reported here as 0.0. (b) t-statistic $t_0$ for all pairwise comparisons. We note that the matrix is symmetric about the diagonal if the $t_0$ value is multiplied by -1. (c) Degrees of freedom $\nu$ for all pairwise comparisons.}
    \label{AppFig: N level stats supplement}
\end{figure}
where $\Gamma$ is the gamma function. 
The t-distribution PDF is a symmetric function about $t= 0$, greatly resembling a bell curve of mean zero and variance one, but typically with more weight in the tails of the distribution. 
As $\nu$ increases, the t-distribution PDF approaches the normal distribution.

The p-value is calculated as the integral of $f(t, \nu)$, from values of $\pm t_0$ to $\pm \infty$.
Graphically, it is the area in the tails of the distribution, defined for the regions above $t_0$ and below $-t_0$.
If a distribution is compared to itself under this test, the t-statistic $t_0 = 0$, and the p-value would therefore be equal to 1.
For two distinct distributions, a small p-value indicates that the observed difference between the distributions is unlikely to be random, indicating that according to the test, their underlying distributions are different.  

It is additionally insightful to calculate the confidence interval (CI) of the t-statistic: the range of values expected to contain the true $t_0$ of the distribution, to a set degree of confidence. 
The CI is calculated as 
\begin{equation}
\pm  t_{\alpha/2, \nu}\times \sqrt{\frac{\sigma_1^2}{n_1} + \frac{\sigma_2^2}{n_2}}
\label{eq: t-test confidence interval}
\end{equation}
where $t_{\alpha/2}$ is the critical t value associated with a set level of confidence $(1-\alpha)\times 100\%$, for a PDF corresponding to $\nu$ degrees of freedom. 
The critical t value is the value at which the associated PDF has the integrated area $\alpha/2$ under its curve in the regions defined as above the value $t_{\alpha/2}$ and below the values -$t_{\alpha/2}$.

\renewcommand{\arraystretch}{1.5}
\begin{table*}[t!h]
\centering
\begin{tabular}{c||c  c  c  c  c  c  c  c } 
 & $\overline{A1}$ & $\overline{A2}$ & $\overline{A3}$ & $\overline{A4}$ & $\overline{A5}$ & $\overline{B1}$ & $\overline{B2}$ & $\overline{B3}$ \\
\hline
$A1$ & 0$\pm$16$\%$ & -13$\pm$17$\%$ & 7$\pm$17$\%$ & 33$\pm$20$\%$ & 100$\pm$26$\%$ & -25$\pm$11$\%$ & 11$\pm$19$\%$ & -18$\pm$14$\%$ \\
\hline
$A2$ & 15$\pm$20$\%$ & 0$\pm$20$\%$ & 23$\pm$21$\%$ & 53$\pm$26$\%$ & 131$\pm$35$\%$ & -14$\pm$14$\%$ & 28$\pm$23$\%$ &  -5$\pm$17$\%$ \\
\hline
$A3$ & -6$\pm$16$\%$ & -19$\pm$17$\%$ & 0$\pm$16$\%$ & 24$\pm$19$\%$ & 87$\pm$24$\%$ & -30$\pm$11$\%$ & 4$\pm$18$\%$ & -23$\pm$13$\%$ \\
\hline
$A4$ &  -25$\pm$15$\%$ & -35$\pm$17$\%$ & -20$\pm$16$\%$ & 0$\pm$19$\%$ & 50$\pm$23$\%$ & -44$\pm$10$\%$ & -17$\pm$18$\%$ & -38$\pm$13$\%$\\
\hline 
$A5$ & -50$\pm$23$\%$ &  -57$\pm$15$\%$ &  -47$\pm$13$\%$ & -34$\pm$16$\%$ & 0$\pm$17$\%$ & -63$\pm$9$\%$ & -45$\pm$16$\%$ & -59$\pm$11$\%$\\
\hline
$B1$ & 34$\pm$15$\%$ & 16$\pm$17$\%$ & 43$\pm$15$\%$ & 78$\pm$19$\%$ & 168$\pm$23$\%$ & 0$\pm$10$\%$ & 48$\pm$18$\%$ & 10$\pm$13$\%$ \\
\hline
$B2$ &  -10$\pm$17$\%$ & -22$\pm$18$\%$ & -3$\pm$18$\%$ & 20$\pm$22$\%$ & 81$\pm$28$\%$ & -33$\pm$12$\%$ & 0$\pm$20$\%$ &  -26$\pm$15$\%$\\
\hline
$B3$ & 21$\pm$17$\%$ & 5$\pm$18$\%$ & 30$\pm$17$\%$ & 61$\pm$21$\%$ & 142$\pm$28$\%$ & -9$\pm$12$\%$ & 34$\pm$20$\%$ & 0$\pm$15$\%$ \\
\end{tabular}
\caption{Confidence interval of 95$\%$ significance level for Welch's t-test pairwise comparison of devices in the main text. Intervals are reported as the percentage of the column qubit's $\qceff$ distribution mean, where negative (positive) values indicate the row qubit mean is lower (higher) than the column qubit mean.
\label{tab: CIs 95}}

\end{table*}

A CI that contains the value zero indicates that to the set level of confidence, the test cannot claim that their means are distinct. 
A CI containing zero will correspond to a p-value above the threshold corresponding the same set confidence level.
A CI that contains only negative or only positive values indicates that the compared sample distribution means are different up to the confidence value of the test.
Negative (positive) values indicate the mean of distribution 1 is less (more) than the mean of distribution 2. 

We report here the test details for pairwise comparisons of all qubits. 
We show the p-values, t-statistic, and degrees of freedom for this comparison in Fig \ref{AppFig: N level stats supplement} and 95$\%$ CIs in Table \ref{tab: CIs 95}. 
CIs for each pairwise comparison are reported as a percent of the mean of the column qubit $\qceff$ distribution. 
We note that we observe CI that are generally wide compared to this mean, indicating that the test cannot easily distinguish between individual qubits.
Taken together, the p-values and CIs indicate that Welch's t-test identifies differences between distributions as small as approximately $20\%$ of a qubit's mean. 
From these features, we conclude that our analysis protocol is sensitive enough to resolve differences at this level, and that the distribution of $\qceff$ between qubits are quite similar. 

We also compare between fabrication processes using the combined distributions of qubits $A1$, $A2$, and $A3$ as one distribution, and the combined distributions of qubits $B1$, $B2$, and $B3$ as the other distribution.
We report the 95$\%$ confidence intervals for these comparisons in Table~\ref{tab: CIs 95 process compare}. 
We additionally report that for this comparison, the p-value was 0.001, the t-statistic $t_0 = -3.22$, and $\nu = 889.6$.
This indicates that the pad etch treatment of Process $B$ only slightly improves the mean $\qceff$ as compared to Process $A$.

\renewcommand{\arraystretch}{1.5}
\begin{table}[t!h]
\centering
\begin{tabular}{c||c | c} 
 & $\overline{A}$ & $\overline{B}$ \\
\hline
$A$ & 0$\pm$10$\%$ & -16.0$\pm$9.7$\%$ \\
\hline
$B$ & 13.8$\pm$8.4$\%$ & 0$\pm$8$\%$ \\
\hline
\end{tabular}
\caption{Confidence interval of 95$\%$ significance level for Welch's t-test pairwise comparison of combined process distributions in the main text. Intervals are reported as the percentage of the column qubit's $\qceff$ distribution mean, where negative (positive) values indicate the row qubit mean is lower (higher) than the column qubit mean.
\label{tab: CIs 95 process compare}}

\end{table}

\section{\texorpdfstring{$\qceff$}{qceff} Dependence on \texorpdfstring{$P_{\mathrm{JJ}}$}{PJJ}}
\label{App: N level JJ Participation Ratio}
Here, we investigate whether there is a trend in the mean $\qceff$ with the area of the small JJ, expressed as its percent contribution to the total qubit capacitance $P_{\mathrm{JJ}}$. 
To calculate $P_{\mathrm{JJ}}$, we assume a junction specific capacitance of 49 $\mathrm{fF}/\mu {\mathrm{m}}^2$ as extracted by Ref.~\cite{randeria2024CQPS} for qubits utilizing the same fabrication process as our process $A$.
\begin{figure}[t!]
    \centering
    \includegraphics{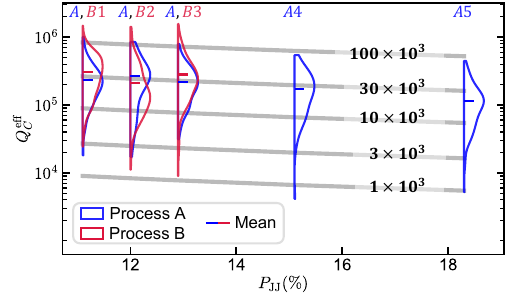}
    \caption{Qubit $\qceff$ distributions versus small JJ participation ratio. The area of each distribution is normalized to show the percent of counts at each $\qceff$. Horizontal lines for different values of $Q_\mathrm{JJ}$ are meant to guide the eye.}
    \label{AppFig: N level JJ Participation Ratio}
\end{figure}
Across designs 1-5, the single junction size of the qubit was varied to give the range in $\EJ$ reported in Table~\ref{tab:Fluxonium Device Summary}. 
The varied range in $\EJ$ corresponds to approximately an 11-18$\%$ range of the single junction contribution to the total qubit circuit capacitance.
We plot the extracted $\qceff$ distributions versus their small JJ capacitance participation ratio in Fig. \ref{AppFig: N level JJ Participation Ratio}. 
Modeling the total qubit quality factor $\qceff$ as
\begin{equation}
    1/\qceff = P_{\mathrm{JJ}}/Q_{\mathrm{JJ}} + (1- P_{\mathrm{JJ}})/Q_{\mathrm{other}},
\end{equation}
where $Q_{\mathrm{JJ}}$ is the materials quality factor of the junction and $Q_{\mathrm{other}}$ the materials quality factor of the rest of the circuit, we map between our measured distributions and the $Q_{\mathrm{JJ}}$, assuming $Q_{\mathrm{other}}\gg Q_\mathrm{JJ}$.
Lines to guide the eye for different values of $Q_{\mathrm{JJ}}$ are included in Fig. \ref{AppFig: N level JJ Participation Ratio}.

We observe a weak trend in the distribution maximum $\qceff$, as well as the distribution mean with the junction capacitance participation ratio for Process $A$. 
This is somewhat expected --- there will be a contribution from the metal-substrate interface that was nominally improved by Process $B$ in this study that scales with JJ size, and other measurements of JJ barrier quality give similar effective $Q_\mathrm{JJ}$~\cite{Mamin_2021}.
We note that qubits $A1, A2$, and $A3$ were fabricated on the same sample, with $A4, A5$ on a second sample, both from the same wafer. 
We do not expect a high variation in performance from sample to sample on the same wafer in our fabrication process. 
We note that this study did not seek to quantify the single-junction branch contribution to the qubit $\qceff$.  

\section{Comparison assuming frequency independent $\qceff$}
\begin{figure}[hb!]
    \centering
    \includegraphics{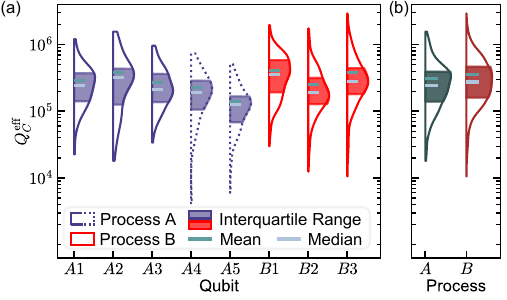}
    \caption{Frequency independent $\qceff$ analysis: qubit $\qceff$ distributions, where under Eq.~\ref{eq: Freq dep Qc}, $\epsilon = 0$. The area of each distribution is normalized to show the percent of counts at each $\qceff$. (a) Individual qubit $\qceff$ distributions. Blue color corresponds to qubits fabricated with process $A$; red corresponds to process $B$. Shaded regions within individual distributions represent the interquartile range of the dataset. Overlaid teal and gray lines mark the mean and median of each distribution, respectively. (b) Combined distributions of qubits $A1$, $A2$, and $A3$ marked as Process $A$ (dark teal), and $B1$, $B2$, and $B3$ (Process $B$, dark red) are used to compare like designs across fabrication processes. Dashed outlines in (a) indicate exclusion from the combined distributions shown in panel (b).}
    \label{AppFig: Comparison under epsilon = 0}
\end{figure}
\label{app: N level compare with 0 epsilon}
\begin{table}[h]
\begin{tabular}{c||c|c|c|c|} 
Qubit & Mean $\qceff$ & Median $\qceff$ & $\sigma$ & 95$\%$ CI\\
\hline
$A1$ &  2.90$\times10^5$ & 2.44$\times 10^5$ &  2.05$\times 10^5$ & $\pm$ 17$\%$\\
\hline
$A2$ & 3.80$\times10^5$ & 3.19$\times 10^5$ & 3.33$\times 10^5$ &  $\pm$ 24$\%$\\
\hline
$A3$ & 2.73$\times 10^5$ & 2.14$\times 10^5$ & 1.95$\times 10^5$ &  $\pm$ 18$\%$\\
\hline
$A4$ & 2.26$\times10^5$ & 1.93$\times 10^5$ & 1.52$\times 10^5$ &  $\pm$ 20$\%$ \\
\hline 
$A5$ & 1.41$\times10^5$  &  1.26$\times 10^5$ & 0.99$\times 10^5$ &  $\pm$ 17$\%$\\
\hline
$B1$ & 4.12$\times10^5$ & 3.58$\times 10^5$  & 2.84$\times 10^5$ & $\pm$ 11$\%$\\
\hline
$B2$ & 2.56$\times10^5$ & 1.91$\times 10^5$  & 2.36$\times 10^5$ & $\pm$ 18$\%$ \\
\hline
$B3$ & 3.78$\times10^5$ & 2.79$\times 10^5$  &  3.49$\times 10^5$ & $\pm$ 18$\%$\\
\hline
$A$ & 3.11$\times10^5$ & 2.44$\times10^5$  &  2.50$\times 10^5$ & $\pm$ 12$\%$\\
\hline
$B$ &  3.61$\times10^5$ & 2.75$\times10^5$  &  2.99$\times 10^5$ & $\pm$ 9$\%$\\
\end{tabular}
\caption{Frequency independent $\qceff$ analysis: $\qceff$ distribution summary for each qubit and for the combined process distributions, with $\epsilon = 0$. Means, medians, and standard deviations $\sigma$ of each qubit distribution are reported. 95$\%$ confidence interval (CI) of the mean as returned by Welch's t-test for each $\qceff$ distribution as compared to itself, expressed as a percent of the distribution mean. This interval demonstrates the sensitivity with which this method can be used to resolve differences between distributions.
\label{tab:epsilon = 0 Level Mean, Median, CIs} }
\end{table}

In the main text and the majority of analysis, we choose to add a phenomenological frequency dependence to the dielectric quality factor introduced in literature~\cite{Wang2015SurfacePartcipationandDielectric, Nguyen_2019, Sun_2023, Ardati_2024} as it more closely matches the trends in our data.
Here, we compare the distributions of qubit $\qceff$ without this frequency dependence.

We perform the identical analysis described in the main text and Appendices \ref{App: Data Processing}, \ref{App: Data Processing: Binning}, \ref{App: Stats Tests}, substituting $\epsilon$ = 0 into~\eqlbl~\ref{eq: Freq dep Qc}.
The means, medians, standard deviations, and self confidence intervals for individual qubit distributions and the combined process distributions are reported in Table~\ref{tab:epsilon = 0 Level Mean, Median, CIs}.
We see generally higher values of each mean, consistent with the understanding that the value of $\epsilon$ used in the main text decreases the $\qceff$ values near $\Phi_\mathrm{ext}/\Phi_0 = 0.5$. 
We apply Welch's t-test to these distributions, calculating the 95$\%$ CI of the distributions as compared to themselves to obtain an average test sensitivity of ±18$\%$ of an individual qubit’s mean [Table~\ref{tab:epsilon = 0 Level Mean, Median, CIs}].

Combining the distributions of $A1, A2, A3$ and $B1, B2, B3$ we compare the two fabrication processes used in this study.
Welch's t-test reports these distributions to be different, with a 95$\%$ confidence interval indicating that mean of distribution of $B$ qubits is higher by $16.2\pm 11.0\%$ of the mean of the corresponding $A$ qubits.
This is consistent with the main text result that process $B$ did reduce the loss from the metal-substrate interface, but that this interface is not the dominant limitation to the performance of these fluxoniums.

\section{Comparison Using a 2-Level Model}
\label{App: 2 level modeling}
The analysis presented in the main text utilizes a six-level model to describe the energy relaxation dynamics in the fluxonium qubits. 
For comparison with literature, here we perform the identical analysis assuming the population dynamics remain solely within the qubit states $|0\rangle$ and $|1\rangle$. 

We again apply the 8 MHz binned average to the dataset, and transform the downselected dataset into a two-level $\qceff$.
As done in the main text, we combine three models to describe the measured $T_1$ versus device transition frequency: capacitive dielectric loss, $1/f$ flux noise, and radiative loss to the microwave environment. 
We invert the following addition of rates,
\begin{equation}
    \Gamma_1^{\mathrm{meas}} = \Gamma_1^{\mathrm{C}} + \Gamma_1^{\Phi} + \Gamma_1^{\mathrm{P}} + \Gamma_1^{\mathrm{cd}} + \Gamma_1^{\mathrm{fd}}
\end{equation}
where $\Gamma_1^{\mathrm{meas}}$ the measured relaxation rate, to solve for $\qceff$ (contained within $ \Gamma_1^{\mathrm{C}}$).
We find under this model that a phenomenological frequency dependence of the quality factor as described by Eq.~\ref{eq: Freq dep Qc} describes the data best.

The number of points in the distributions before and after the binning is applied is the same as in the case of the N-level analysis.
We summarize the number of points at each analysis step in Table~\ref{tab:2, N level qubit T1 sample sizes}.
A main difference in this analysis from that in the main text is the shape and magnitude of the limitation due to Purcell loss through the readout resonator, though we see no difference in the number of points included between our two model descriptions [Fig.~\ref{AppFig: All T1s, coherence model overlay}]. 

Points for which the predicted rate from the combination of the two-level models for flux noise and radiative losses is greater than 10$\%$ of the measured rate were removed from this analysis.
All remaining data is assumed to be limited by capacitive dielectric loss.
We find that $\epsilon = 0.2$ best describes the extracted $\qceff$ for all qubits, which is slightly different than the $\epsilon = 0.25$ that fits the dataset under the six-level analysis.

\renewcommand{\arraystretch}{1.5}
\begin{table}[h]
\centering
\begin{tabular}{c||c|c|c|c|} 
Qubit & $\#T_1$ & $\#$Post-binning & $\#$6-level $\qceff$ & $\#$2-level $\qceff$\\
\hline\hline
$A1$ & 209 & 144 & 137 & 137 \\
\hline
$B1$ & 530 & 314 & 314 & 314\\
\hline
$A2$ & 155 & 107 &  104 & 104\\
\hline
$B2$ & 398 & 188 & 188 & 188\\
\hline
$A3$ & 257 & 120  & 118 &  119 \\
\hline
$B3$ & 427 & 213 & 211 & 211\\
\hline
$A4$ & 221 &  102 & 93 & 93 \\
\hline
$A5$ & 270 & 135 & 130 & 130\\
\hline
\end{tabular}
\caption{Total number of $T_1$ measurements taken across the flux range of each qubit, number of $T_1$ points after 8 MHz binned average, number of points in each extracted multilevel $\qceff$ distribution, and in the two-level $\qceff$ distributions. Fewer points in the $\qceff$ column than the ``Post-binning" column indicate that some points were limited by the models used for $1/f$ flux noise or spontaneous emission to the control lines. Differences in the six-level $\qceff$ and two-level $\qceff$ indicate that the model description is important for classifying whether points are limited by the dielectric loss model, according to the imposed rate requirements detailed in the text.
\label{tab:2, N level qubit T1 sample sizes}
}
\end{table}

\begin{table}[h]
\begin{tabular}{c||c|c|c|c|} 
Qubit & Mean $\qceff$ & Median $\qceff$ & $\sigma$ & 95$\%$ CI\\
\hline
$A1$ &  2.23$\times10^5$ & 1.97$\times 10^5$ &  1.43$\times 10^5$ & $\pm$ 15$\%$\\
\hline
$A2$ & 2.66$\times10^5$ & 2.39$\times 10^5$ & 1.98$\times 10^5$ &  $\pm$ 20$\%$\\
\hline
$A3$ & 2.00$\times 10^5$ & 1.83$\times 10^5$ & 1.22$\times 10^5$ &  $\pm$ 16$\%$\\
\hline
$A4$ & 1.58$\times10^5$ & 1.28$\times 10^5$ & 1.11$\times 10^5$ &  $\pm$ 20$\%$ \\
\hline 
$A5$ & 0.90$\times10^5$  &  0.75$\times 10^5$ & 0.65$\times 10^5$ &  $\pm$ 18$\%$\\
\hline
$B1$ & 3.00$\times10^5$ & 2.54$\times 10^5$  & 2.05$\times 10^5$ & $\pm$ 11$\%$\\
\hline
$B2$ & 2.07$\times10^5$ & 1.35$\times 10^5$  & 2.05$\times 10^5$ & $\pm$ 20$\%$ \\
\hline
$B3$ & 2.67$\times10^5$ & 2.17$\times 10^5$  &  2.10$\times 10^5$ & $\pm$ 15$\%$\\
\hline
$A$ & 2.28$\times10^5$ & 2.00$\times10^5$  &  1.57$\times 10^5$ & $\pm$ 10$\%$\\
\hline
$B$ &  2.66$\times10^5$ & 2.09$\times10^5$  &  2.10$\times 10^5$ & $\pm$ 10$\%$\\
\end{tabular}
\caption{Two-level analysis: $\qceff$ distribution summary for each qubit and for the combined process distributions. Means, medians, and standard deviations $\sigma$ of each qubit distribution are reported. 95$\%$ confidence interval (CI) of the mean as returned by Welch's t-test for each $\qceff$ distribution as compared to itself, expressed as a percent of the distribution mean. This interval demonstrates the sensitivity with which this method can be used to resolve differences between distributions.
\label{tab:2 Level Mean, Median, CIs} 
}
\end{table}

\begin{figure}[t]
    \centering
    \includegraphics{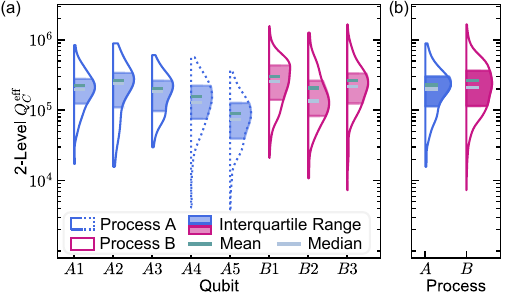}
    \caption{Two-level analysis: Comparison of the extracted $\qceff$ distributions. The area of each distribution is normalized to show the percent of counts at each $\qceff$. (a) All individual qubit $\qceff$ distributions. Blue distributions correspond to qubits fabricated with process A, pink distributions correspond to process B. Shaded regions within device distributions represent the interquartile range of the dataset, with overlaid teal and gray lines marking the mean and median of each distribution, respectively. (b) Combined distributions of qubits $A1$, $A2$, and $A3$, marked as Process $A$ (dark blue), and $B1$, $B2$, and $B3$ (Process $B$, dark pink) are used to compare like designs across fabrication processes. Dashed outlines in (a) indicate exclusion from the combined distributions shown in panel (b).}
    \label{AppFig: 2 level device comparison}
\end{figure}

The distributions are then compared to each other using Welch's t-test (Fig. \ref{AppFig: 2 level device comparison}).
We observe similar results to the main text under this two-level analysis. 
The extracted $\qceff$ distributions are again largely overlapped and non-Gaussian distributed, with a measured range of similar width. 
We calculate the 95$\%$ self-CI of the distributions and obtain an average test sensitivity of ±17$\%$ of an individual qubit’s mean [Table~\ref{tab:2 Level Mean, Median, CIs}].
Comparing designs within process $A$, we observe a similar, weak trend in the $\qceff$ mean with $\EJ$ as shown for the six-level model in Appendix~\ref{App: N level JJ Participation Ratio}.

We combine the distributions of $A1, A2$, and $A3$ and those of $B1, B2$, and $B3$ to compare the like designs across fabrication processes.
Welch's t-test finds these two composite distributions to be distinct, with a 95$\%$ confidence interval that the mean of distribution of $B$ qubits is higher by $16.6\pm 9.8\%$ of the mean of the corresponding $A$ qubits.
Under this analysis, we again see that the process $B$ did reduce the loss from the metal-substrate interface as compared to process $A$.
This small improvement indicates that the metal-substrate interface is not dominating the dissipation in these fluxonium.
Though this analysis utilizes a less well motivated description of the decay dynamics than in the main text, we obtain results consistent with the main text conclusions. 
\clearpage
\bibliography{bibliography.bib}
\end{document}